\documentclass[preprint,3p]{elsarticle}
\usepackage[utf8]{inputenc}

\usepackage{amsfonts, amssymb, amsmath}
\usepackage{textgreek}
\usepackage{float}
\usepackage[dvipsnames]{xcolor}
\newcommand{\uu}{{\bf u}}
\newcommand{\code}{\texttt}
\newcommand{\cc}{\color{black}} 
\newcommand{\ccr}{\color{black}} 

\title{ALLIANCE: Spectral solver for kinetic plasma turbulence}

\date{}

\begin{document}

\author[1]{Evgeny A. Gorbunov\corref{cor1}%
}
\ead{gorbunove@uni.coventry.ac.uk}
\affiliation[1]{organization={Coventry University},
addressline={Priory street},
postcode={CV1 5FB},
city={Coventry},
country={United Kingdom}}

\author[2]{Bogdan Teaca}\corref{cor2}
\ead{bteaca@gmail.com}
\affiliation[2]{organization={University of Craiova},
addressline={3 A.I. Cuza Street},
city={Craiova},
postcode={200585},
country={Romania}}

\begin{abstract}
The \code{ALLIANCE}\footnote{{\cc ALLIANCE - spectrAL soLver for kInetic plAsma turbuleNCE}} code is developed to solve a new set of four-dimensional electromagnetic drift-kinetic equations in slab geometry \cite{gorbunov_2022}. The nonlinear equations are useful for the study of magnetized plasma systems at scales comparable to, or larger than the ion gyroradius. In particular, it is suited for the study the kinetic turbulent cascade in astrophysical plasma, while preserving finite Larmor radius effects at the fluid-kinetic transition. The equations solved are in spectral Fourier-Laguerre-Hermite form, a pseudo-spectral approach is used for the nonlinear terms, and the code is parallelised over multiple directions. After a presentation of the code, validation runs are shown, and benchmarks for serial and parallel computations are presented.  
\end{abstract}


\maketitle

\section{Introduction}
In strongly magnetized plasma, gyrokinetic (GK) formalisms \cite{brizard_2007} are employed for the study of fusion in laboratory \cite{Fasoli_2016, Helander_2015} and kinetic Alfv\'en wave turbulence in astrophysical conditions \cite{howes_2006, chen_2013, teaca_2016}. Obtained from a Maxwell-Vlasov system of kinetic equations \cite{eyink_2018}, one of the basic assumptions in deriving the gyrokinetic formalism is that the fast gyration of the charged particles can be integrated out of the equations of motions, thus reducing the dimensionality of the problem from six to five dimensions. The impact of polarization effects due to the fast gyromotion are taken into account via gyroaveraging operators, which in Fourier space are represented by a simple product with Bessel functions. The physical meaning behind the gyroaveraging operation is to take into account finite Larmor radius (FLR) of the gyrating particles. 

GK formalisms have been effectively used for the study of astrophysical plasma turbulence at scales ($\ell \sim 1/k$) of the size of the ion gyroradius (i.e. $k_\perp \rho_i\approx 1$) and smaller. At these scales, the plasma is strictly kinetic and fluid descriptions are insufficient to capture the relevant linear and nonlinear phase mixing. However, since the nonlinear mixing of velocity structures becomes weak at scales above the ion gyroradius, GK equations are becoming computationally expensive considering the relevant physics being solved. Thus, simplified models have been proposed \cite{zocco_2011, hatch_2014, kunz_2015, brizard_1992} for scales much larger than the ion gyroradius ($k_\perp \rho_i\gg 1$) in an attempt to understand the effective fluid-kinetic system in astrophysical conditions. Such approach is helped by the fact that the ratio $l_\parallel/l_\perp$ between  parallel and perpendicular characteristic scales for solar wind, for example, is large \cite{chen_2016} (with $l_\parallel \approx 1$ AU, and $l_\perp = \rho_i \approx 10^2$ km \cite{Kiyani_2015}). Moreover, with in-situ solar wind observations \cite{alexandrova_2008, alexandrova_2009, duan_2020} and numerical simulations \cite{tatsuno_2009, told_2015,meyrand_2021} showing the existence of a so-called spectral "knee", characterized by a spectral break for different quantities at the $k_\perp \rho_i\approx 1$ scale, a simplified set of equations that account for this transition at least qualitatively is also desired. 

The exact mechanism for the ion Larmor radius transition is still under scrutiny, and requires a model that can allows us to study this and the fluid-kinetic transition in detail, while being more computationally effective compared to GK. In order to study the scales above the $k_\perp \rho_i \approx 1$ in a meaningful way and with as little limiting assumptions as possible, one can employ a Fourier-Laguerre-Hermite decomposition \cite{mandell_2018}, keep the relevant FLR dynamics that become relevant at $k_\perp \rho_i \approx 1$ by taking into account by the first two (0th and 1st) Laguerre moments of the gyrokinetic distribution function, and only then apply the drift-kinetic limit $k_\perp \rho_i \ll  1$. The resulting set of electromagnetic drift-kinetic equations are four-dimensional, with three spatial directions and a parallel velocity one. This new set of four-dimensional electromagnetic drift-kinetic equations in slab geometry that accounts for FLR effects was derived by the authors in \cite{gorbunov_2022}. 

In order to solve this drift-kinetic system, the new code \code{ALLIANCE} was developed and is presented here. \code{ALLIANCE} evolves in time the first two Laguerre moments of the GK distribution function, with the nonlinear terms being solved using a pseudo-spectral approach for the spatial directions, and the parallel velocity direction being captured via a Hermite decomposition. Such spectral formulation allows for to study of kinetic turbulent cascade in the spatial directions while accounting for parallel velocity mixing, therefore providing a link between fluid and kinetic turbulence. 

The paper is organized as follows: in section \ref{sec:equations}, we introduce the drift-kinetic equations in the form solved numerically, mentioning all the necessary preliminary information needed to give the reader a clear understanding of the physical system. Next, in section \ref{sec:numericalImplementation}, we talk about the numerical implementation of the equations, such as parallelization and numerical schemes. In section \ref{sec:CodeValidation} we present initial simulation results, and several numerical tests to validate the code. Single processor and multiprocessor benchmarks are given in \ref{sec:PerformanceRuns}. Last, we discuss the results, and propose the possible usage of the code, as well as further developments in section \ref{sec:Conclusion}.

\section{Drift-kinetic model of plasma}\label{sec:equations}
\begin{figure}[b]
\centering 
\includegraphics[width=1.0\textwidth]{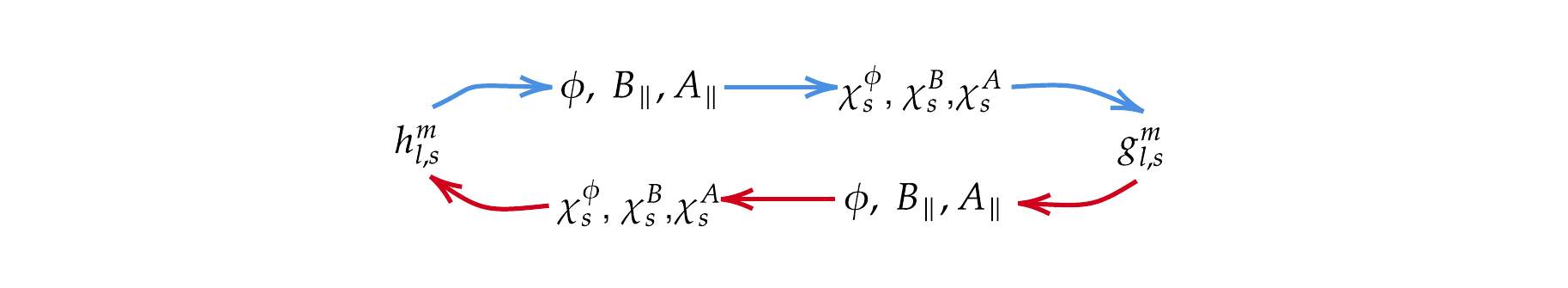}
\caption{Transition from $g^m_{l,s}$ to $h^m_{l,s}$. First, the electromagnetic fields are computed using $g^m_{ls}$ (respectively $h^m_{l,s}$), then the GK potentials are computed from the fields, and finally the relation \eqref{eq:g} is used to compute $h^m_{l,s}$ (respectively $g^m_{l,s}$). An initial condition can be given in term of $g^m_{l,s}$ or $h^m_{l,s}$, however, the equations are integrated in time in term of $g^m_{l,s}$.}
\label{fig:gToH}
\end{figure}

In this section, we present the equations which are solved by \code{ALLIANCE}. Here, we do not provide the comprehensive derivation of the equations, which can be found in \cite{gorbunov_2022}, and list only the minimum information required for understanding the model used. First, we consider a plasma with particles of species $s$, influenced by a straight strong magnetic guide field, acting in the $z$-direction, $\mathbf{B} = B_0 \mathbf{\hat{z}}$. For a GK system, the particle perturbed distribution function for species $s$ can be found as $\delta f_s= - q_s \phi F_s/T_s + h_s$, with $h_s(\mathbf{R}, v_\parallel,\mu)$ being the non-adiabatic part of the GK distribution function. 
Using a spectral formulation \cite{mandell_2018}, it can be represented by its Laguerre-Hermite moments $h^m_{l,s}$ as  
$h_s(\mathbf{k},v_\parallel,\mu) = \sum^{\infty}_{m=0}\sum^{\infty}_{l=0}\psi^l\left(\mu B_0\right)\xi^m\left(v_\parallel\right) h^m_{l,s}(\mathbf{k})$
where $\psi^l $, and $\xi^m $ are Laguerre and Hermite functions, respectively \cite{mandell_2018}. For a drift-kinetic tendency, seen here as $k_\perp \rho_i < 1$ rather than the stringent limit condition $k_\perp \rho_i \ll 1$, all the high Laguerre moments $l\ge 2$ can be omitted as negligibly small. Such simplification preserve finite Larmor radius (FLR) effects, while significantly simplifying the gyrokinetic system \cite{gorbunov_2022}. 
The gyrokinetic distribution function moments are related to the gyrocenter distribution function moments via the relation 
\begin{equation}\label{eq:g}
    g^m_{l,s}(\mathbf{k}) = h^m_{l,s}(\mathbf{k}) - \frac{q_s}{T_s}(\chi_s^{\phi} +\chi_s^{B})\delta_{m0}\delta_{l0} - \frac{q_s}{T_s}\sqrt{\frac{1}{2}}\chi_s^{A} \delta_{m1}\delta_{l0} - \frac{q_s}{T_s} \chi_s^{B}\delta_{m0}\delta_{l1}.
\end{equation}
The difference between the $g_s$ and $h_s$ can be seen as following: the $g_s$ describes the point-like particle gyrocentres, while $h_s$ is the distribution function of the charged rings (see C.5 in \cite{kunz_2015}). Relation \eqref{eq:g} involve gyrokinetic potential functions, 
\begin{align}
    \chi^{\phi}_s(\mathbf{k}) &= \mathcal{J}_{s00}\phi(\mathbf{k}) \,,\label{eq:chiPhi}\\
    \chi^{B}_s(\mathbf{k}) &=  \frac{T}{q B_0} \tilde{\mathcal{J}}_{s10}B_\parallel(\mathbf{k}) \,,\label{eq:chiB}\\
    \chi^{A}_s(\mathbf{k}) &= - v_{T_s} \mathcal{J}_{s00} A_\parallel(\mathbf{k})\,,\label{eq:chiA}
\end{align}
which are computed via the electromagnetic fields. {\cc Electromagnetic fields are represented by electrostatic potential $\phi(\mathbf{k})$, 
parallel magnetic field 
$B_\parallel(\mathbf{k})$
and parallel component of the vector potential 
$A_\parallel(\mathbf{k})$ given as:
\begin{align}
    &\phi(\mathbf{k})=\sum_s  q_s n_s \mathcal{J}_{s00}h^0_{0,s} \bigg{/} \sum_s\frac{q_s^2 n_s}{T_s},\\
    &B_\parallel(\mathbf{k})=-\frac{\beta}{2}\sum_s \frac{n_s T_s}{B_0} \tilde{\mathcal{J}}_{s10} \left(h^0_{0,s} +h^0_{1,s}\right),\\
    &A_\parallel(\mathbf{k}) = \frac{\beta}{2 k_\perp^2}\sum_s  q_s n_s v_{T_s} \sqrt{\frac{1}{2}} \mathcal{J}_{s00}h^1_{0,s}. \label{eq:A}
\end{align}}
In terms of $g^m_{l,s}$ the equations for the electromagnetic fields becoming more cumbersome, 
\begin{align}
    &\phi(\mathbf{k}) = \frac{2 b I_\phi(g)  - \beta c I_{B}(g)}{2 ab + \beta c^2},\label{eq:numerics:phi} \\
    &B_\parallel(\mathbf{k}) = -\beta\frac{a I_{B}(g) + c I_\phi(g)}{2 ab +\beta c^2}.\label{eq:numerics:B}
\end{align}
Here, we introduced a number of parameters for ease of notation,
\begin{align}\label{eq:numerics:coefficients}
        & I_\phi(g) = \sum_s q_s n_s \mathcal{J}_{s00} g_{0,s}^0, \\
        & I_B(g) = \sum_s \frac{n_s T_s}{B_0} \tilde{\mathcal{J}}_{s10} \left(g_{0,s}^0 + g_{1,s}^0\right),\\
        & a = {\sum_s \frac{q_s^2 n_s}{T_s}}(1 - {\mathcal{J}_{s00}^2}), \\
        & b = 1+\beta\sum_s \frac{n_s T_s}{B_0^2}\mathcal{\tilde{J}}_{s10}^2, \\
        & c = \sum_s\frac{q_s n_s}{B_0} \mathcal{\tilde{J}}_{s10}\mathcal{J}_{s00}.
\end{align}
The relation for $A_\parallel$ is also modified, and now expressed via $g_{0,s}^1$:
\begin{equation}\label{eq:numerics:A}
    A_\parallel(\mathbf{k}) = \frac{\beta}{2} \frac{\sum_s q_s n_s v_{Ts}\mathcal{J}_{s00}\sqrt{\frac{1}{2}}g_{0,s}^1}{ k_\perp^2+\frac{\beta}{4}\sum_s \frac{q_s^2 n_s v_{Ts}^2}{T_s}\mathcal{J}_{s00}^2}\,.
\end{equation}
The FLR effects in eqs.\eqref{eq:chiPhi}-\eqref{eq:numerics:A} are taken into account via the Bessel functions moments,   
\begin{align}
    &\mathcal{J}_{s00} = e^{-b_s/2},\label{equation:J round 0}\\
    &\mathcal{\tilde{J}}_{s10} = \left[\left(1-e^{-b_s/2}\right)\frac{2}{b_s}\right],\label{equation:J round 1}
\end{align}
{\cc The tilde notation in eq.\eqref{equation:J round 1} is used in order to make a clear distinction between the zeroth Laguerre moment of a first order Bessel function $J_0(\sqrt{2\mu B_0 b_s}) = \sum_l \psi_l \mathcal{J}_{s0l}$ and the zeroth Laguerre moment of the function $2 J_1(\sqrt{2\mu B_0 b_s}) / \sqrt{2\mu B_0 b_s} = \sum_l \psi_l \mathcal{\tilde{J}}_{s1l}$ (see \cite{gorbunov_2022} for details).} 
The Bessel function moments \eqref{equation:J round 0},\eqref{equation:J round 1} depend on the product $b_s = (k_\perp \rho_s)^2$, with $\rho_s$ being the gyration radius of the particle of the species $s$. Now, using the relations eqs.\eqref{eq:g}-\eqref{equation:J round 1} it is possible to compute $h^m_{l,s}$ ($g^m_{l,s}$) from $g^m_{l,s}$ ($h^m_{l,s}$) in self-consistent manner, as shown on scheme \ref{fig:gToH}. 

Having established the connection between the $g^m_{l,s}$ and $h^m_{l,s}$ moments, we can now introduce the evolution equations in the drift-kinetic limit. While it is possible to propagate in time the moments of gyrokinetic distribution function $h^m_{l,s}$, it is more numerically convenient to solve the equations for $g^m_{l,s}$ - a preferred choice in other codes relying on the GK expansion \cite{numata_2010}. The evolution of the $g^m_{l,s}(\mathbf{k})$ in the drift kinetic limit with $m = \overline{0,M}$ Hermite moments, and $l={\{0,1\}}$ Laguerre moments is governed by the $M\times2$ system of equations
\begin{align}
\frac{\partial{{g}}^m_{0,s}}{\partial t} &=\mathcal{N}_{0,s}^m[h] + \mathcal{L}_{0,s}^m[h] + \mathcal{C}_{0,s}^m[g]\label{eq:Vlasov0Hermite}\\
\frac{\partial g^m_{1,s}}{\partial t} &=\mathcal{N}_{1,s}^m[h] + \mathcal{L}_{1,s}^m[h] + \mathcal{C}_{1,s}^m[g]\label{eq:Vlasov1Hermite}
\end{align}
Linear terms describe the local coupling of Hermite moments of $h^m_{l,s}$, which include thermal velocities $v_{Ts} = \sqrt{2 T_s/m_s}$:
\begin{align}\label{eq:linear}
    L_{l,s}^m[h] = - i k_z v_{T_s}\left(\sqrt{\frac{m+1}{2}}h^{m+1}_{l,s} + \sqrt{\frac{m}{2}}h^{m-1}_{l,s}\right) ,
\end{align}
and are dictating parallel dynamics along the unperturbed magnetic guide field $k_\parallel = k_z$,  as well as responsible for development of linear phase mixing in $v_\parallel$ direction \cite{schekochihin_2016}.
Nonlinear interactions have the following form,
\begin{align}
    &\mathcal{N}_{0,s}^m[h] = - \frac{1}{B_0}\Bigg[\left\{ \chi_s^{\phi}+\chi_s^{B}, h^m_{0,s}\right\} + \sqrt{\frac{m+1}{2}}\left\{\chi_s^{A} ,h_{0,s}^{m+1}\right\}+\sqrt{\frac{m}{2}}\left\{\chi_s^{A} ,h^{m-1}_{0,s}\right\}+\left\{ \chi_s^{B}, h^m_{1,s}\right\} \Bigg],\label{eq:nonlinearTerm0}\\
    &\mathcal{N}_{1,s}^m[h] = - \frac{1}{B_0}\Bigg[\left\{ \chi_s^{\phi}+\chi_s^{B}, h^m_{1,s}\right\} + \sqrt{\frac{m+1}{2}}\left\{\chi_s^{A} ,h_{1,s}^{m+1}\right\} +\sqrt{\frac{m}{2}}\left\{\chi_s^{A} ,h^{m-1}_{1,s}\right\}+\left\{ \chi_s^{B}, h^m_{0,s}\right\}+2\left\{ \chi_s^{B}, h^m_{1,s}\right\} \Bigg],\label{eq:nonlinearTerm1}
\end{align}
and are expressed in terms of Poisson brackets,  
\begin{align}
    \left\{\mathcal{A},\mathcal{B}\right\} = \frac{\partial \mathcal{A}}{\partial x}\frac{\partial \mathcal{B}}{\partial y} - \frac{\partial \mathcal{A}}{\partial y}\frac{\partial \mathcal{B}}{\partial x},
\end{align}
describing the perpendicular dynamics of the system, $\mathbf{k}_\perp = \{k_x,k_y\}$. Here, nonlinear interactions are mediated by the electromagnetic fields, which enter the entering nonlinear terms in form of the gyrokinetic potentials.

Lastly, \code{ALLIANCE} incorporates a simple collisional term, which has the only purpose to remove energy from small scales. It's form is given as 
\begin{align}\label{eq:collisions}
    \mathcal{C}_{l,s}^m[g] = - \left(\nu_{k_\perp} k^{2\gamma_{k_\perp}}_\perp + \nu_{k_\parallel} k^{2\gamma_{k_\parallel}}_z +\nu_m m^{\gamma_m}\right)g^m_{l,s}(\mathbf{k}).
\end{align}
Collision frequencies in $k_\perp, k_\parallel$ and $m$ directions are set by $\nu_{k_\perp}, \nu_{k_\parallel}, \nu_{m}$. Parameters $\gamma_{k_\perp},\gamma_{k_\parallel},\gamma_{m}$ are provided by the user and are used to define the localization of the dissipation range, and $k_\perp = \left(k_x^2+k_y^2\right)^{1/2}$. {\ccr The chosen collision operator (or similar) has been used before in works \cite{parker_2015,Meyrand_2019,mandell2022gx}. It should be noted that when $\gamma_m = 1$, the collisions in Hermite space become equivalent to Lenard-Bernstein collision operator. As has been shown in \cite{jorge_2018}, such collision operator can impact the dynamics of the system significantly. In future works, other collision operators can be implemented as needed, such as in works \cite{jorge_2017, perrone_2020,frei2022momentbased}, or anomalous dissipation models (i.e. sub-grid-scale models for the nonlinear terms) can be considered \cite{navarro_2014,teaca_2014}.} 

{\ccr It should be noted that since collisional operators can introduce velocity space mixing effects depending on their explicit form, for finite collisional frequency cases in particular (i.e. collisional terms become comparable in intensity with the other linear and nonlinear terms), one must take care that any such model (e.g. \cite{jorge_2017, perrone_2020,frei2022momentbased}) is consistent with the approximations used to derive the overall set of drift-kinetic equations. If collisional terms deemed necessary for a particular project violate the approximations made , a GK or full kinetic formalism may be needed \cite{frei_2022} and {\tt ALLIANCE} is not the suitable tool in that case.}

\section{Numerical implementation}\label{sec:numericalImplementation}
We present next the numerical implementation of our drift-kinetic system. The gyrokinetic distribution function as well as the gyrocenter distribution function moments are represented in memory by 6-dimensional arrays, $g(k_x,k_y,k_z,m,l,s)$ and $h(k_x,k_y,k_z,m,l,s)$. Note that in addition to the three wave-space coordinates $(k_x,k_y,k_z)$ and the Hermite coordinate ($m$), which can all be arbitrary large, the Laguerre coordinate is hard-coded to at most the two elements (i.e. $l=0$ and $l=1$) needed by the drift-kinetic model. Moreover, the species index $s$ track usually the electrons and a single ion species plasma. The data management of the arrays is done with these aspects in mind. The electromagnetic fields are 3-dimensional, with only wave space coordinates $(k_x,k_y,k_z)$, and the gyrokinetic potentials $\chi(k_x,k_y,k_z,s)$ are 4-dimensional. The simulation data is stored on the disk using Hierarchical Data Format (HDF)\cite{hdf5}. 

The choice of space discretization is discussed in the next subsection. Periodic boundaries are assumed for the spatial direction, resulting in a wave-space representation, while a Hermite polynomial representation is used to discretize the parallel velocity direction.
\subsection{Space discretization}
Given a simulation spatial box boundaries $\{L_x,L_y,L_z\}$ and a spatial resolution $\{N_x,N_y,N_z\}$, the wave space discretized with smallest wave numbers $
\{\Delta k_x, \Delta k_y, \Delta k_z\} = \{\frac{2\pi}{L_x},\frac{2\pi}{L_y},\frac{2\pi}{L_z}\}$. The $i$-th wave number is then computed as 
\begin{align}
    &k^i_x = i_x\Delta k_x,\, i_x = \left(-\frac{N_x}{2}+1, ...,\frac{N_x}{2}\right)\,,\\
    &k^i_y = i_y\Delta k_y,\, i_y = \left(-\frac{N_y}{2}+1, ...,\frac{N_y}{2}\right)\,,\\
    &k^i_z = i_z\Delta k_z,\, i_z = \left(0, ...,\frac{N_z}{2}\right)\,.
\end{align}
 A reality condition $f(\mathbf{k}) = f^*(-\mathbf{k})$ is taken into account in $k_z$ direction, allowing to perform computations only for the data above $k_z\ge 0$ plane. The real discrete forward and inverse Fourier transforms for a function $f(\mathbf{r})$ are
\begin{align}
& \widehat{f}(\mathbf{k}) = \sum_\mathbf{r} f(\mathbf{r}) e^{- i\mathbf{k}\cdot\mathbf{r}}\,,\\
& f(\mathbf{r}) = \frac{1}{N_x N_y N_z}\sum_{\mathbf{k}} \widehat{f}(\mathbf{k}) e^{i\mathbf{k}\cdot\mathbf{r}}.
\end{align}
\subsubsection{Fast Fourier Transforms}
The nonlinear terms \eqref{eq:nonlinearTerm0},\eqref{eq:nonlinearTerm1} are computed in real space, which makes it necessary to perform $(2\times N_m \times N_s)$ fast Fourier transforms of size $(N_x \times N_y \times N_z/2)$. 
The nonlinear terms are dealiased using the 2/3 rule, which must be taken into account when initializing the system size. {\cc That is, the nonlinear term computed in real space on the grid of size $(N_x, N_y, N_z)$. Dealiasing is then performed in $(k_x,k_y,k_z)$ space, with $k_{x_{max}}=2\pi N_x/3L_x, k_{y_{max}}=2\pi N_y/3L_y, k_{z_{max}}=2\pi N_z/3L_z$.} FFTs are treated by parallel \code{FFTW} library \cite{FFTW05}, using \code{fft\_mpi\_many} real FFT routine. In order to save the device memory and the computational time, the transforms are performed in-place, and give transposed ($(y,x,z)$ instead of $(x,y,z)$) real output. FFTs in \code{ALLIANCE} are implemented for 6D distribution function moments, electromagnetic fields and gyrokinetic potentials. Another application of the FFTs in \code{ALLIANCE} is to save the real data on the disk.  
\subsection{Time integration}
\subsubsection{RK4}
Equations \eqref{eq:Vlasov0Hermite}-\eqref{eq:Vlasov1Hermite} are propagated in time using the simple order 4  explicit Runge-Kutta scheme. Denoting the RHS of \eqref{eq:Vlasov0Hermite}-\eqref{eq:Vlasov1Hermite} as $\mathcal{R}^m_{l}[g(\tau),h(\tau)]$ at the time $\tau$, we remind the reader the generic algorithm: 
\begin{align}
    &K_1 = \mathcal{R}^m_l[g(\tau),h(\tau)] \,,\\
    &K_2 = \mathcal{R}^m_l\left[g(\tau) + \Delta t\frac{K_1}{2},h(\tau) + \Delta t\frac{K_1}{2}\right]\,,\\
    &K_3 = \mathcal{R}^m_l\left[g(\tau) + \Delta t\frac{K_2}{2},h(\tau) + \Delta t\frac{K_2}{2}\right]\,,\\
    &K_4 = \mathcal{R}^m_l\left[g(\tau) + \Delta t K_3,h(\tau) + \Delta t K_3\right]\,,\\
    &g(\tau+\Delta t ) = g(\tau) + \frac{1}{6}\left(K_1 + K_2 + K_3 + K_4\right)\,,\\
    &\tau = \tau + \Delta t\,.
\end{align}
In case of the linear simulations, size of the time step $\Delta t$ remains constant throughout the run. The necessary requirement for the simulation to be stable and converge, which means that all the eigenvalues of the linear operator \eqref{eq:linear} and, if present, eigenvalues of the collision operator \eqref{eq:collisions} must be localized inside the stability region of the numerical method, i.e. $|1+\lambda \Delta t +\frac{1}{2}\left(\lambda \Delta t\right)^2 + \frac{1}{6}(\lambda \Delta t)^3 + \frac{1}{24}(\lambda \Delta t)^4|\le 1$, where $\lambda$ is an eigenvalue of the linear or collision operators. An a-priory eigenvalue analysis can be performed by the user. 
{\cc One of the possible ways to estimate the linear time step constraint, one can use Gershgorin circle theorem. For the linear operator given by eq.\eqref{eq:linear}, the maximum absolute value of an eigenvalue $\lambda_{max}$ in the system can be estimated as $|\lambda_{max}| \le \max{(k_{z})} \max{(v_{Ts})} \left[\sqrt{(N_m+1)/2}+\sqrt{N_m/2}\right] $. The latter can be used to derive $\Delta t$ from the stability region constraint of the chosen numerical method.}
\subsubsection{Adaptive time stepping}
In case of nonlinear simulations, the situation is more complex, and the usage of the adaptive time step size is required. In this case, we adapt Courant–Friedrichs–Lewy condition, widely used for the nonlinear simulations. In order to use the CFL condition, we first rewrite the Poisson brackets in the form 
\begin{align}
\{A,B\} = \hat{e}_z \times \nabla A \cdot \nabla B,
\end{align}
so we can introduce the following advective velocities:
\begin{align}
    &\uu^{A} = \frac{1}{B_0}\hat{e}_z \times \nabla \chi^A,\\
    &\uu^\phi = \frac{1}{B_0}\hat{e}_z \times \nabla \chi^\phi,\\
    &\uu^B = \frac{1}{B_0}\hat{e}_z \times \nabla \chi^B,
\end{align}
The CFL condition is now can be written as 
\begin{equation}
    \Delta t = \frac{C}{ \max(|\uu_x|)/\Delta x + \max(|\uu_y|)/\Delta y},
\end{equation}
with the Courant number $C = 0.5$ and the maximum drift velocity chosen as 
\begin{equation}
    |\uu| = \max\left\{|\uu^{\phi}|,|\uu^{A}|,|\uu^{B}|\right\}.
\end{equation}
{\cc Depending on the physical parameters, nonlinear or linear time step constraint can dominate. However, for the solar wind parameters at 1 AU ($\beta = T_i/T_e = 1$) for proton-electron plasma, one can expect the linear time step constraint to dominate.}
\subsection{Data parallelization}
\begin{figure}[b]
\centering 
\includegraphics[width=0.75\textwidth]{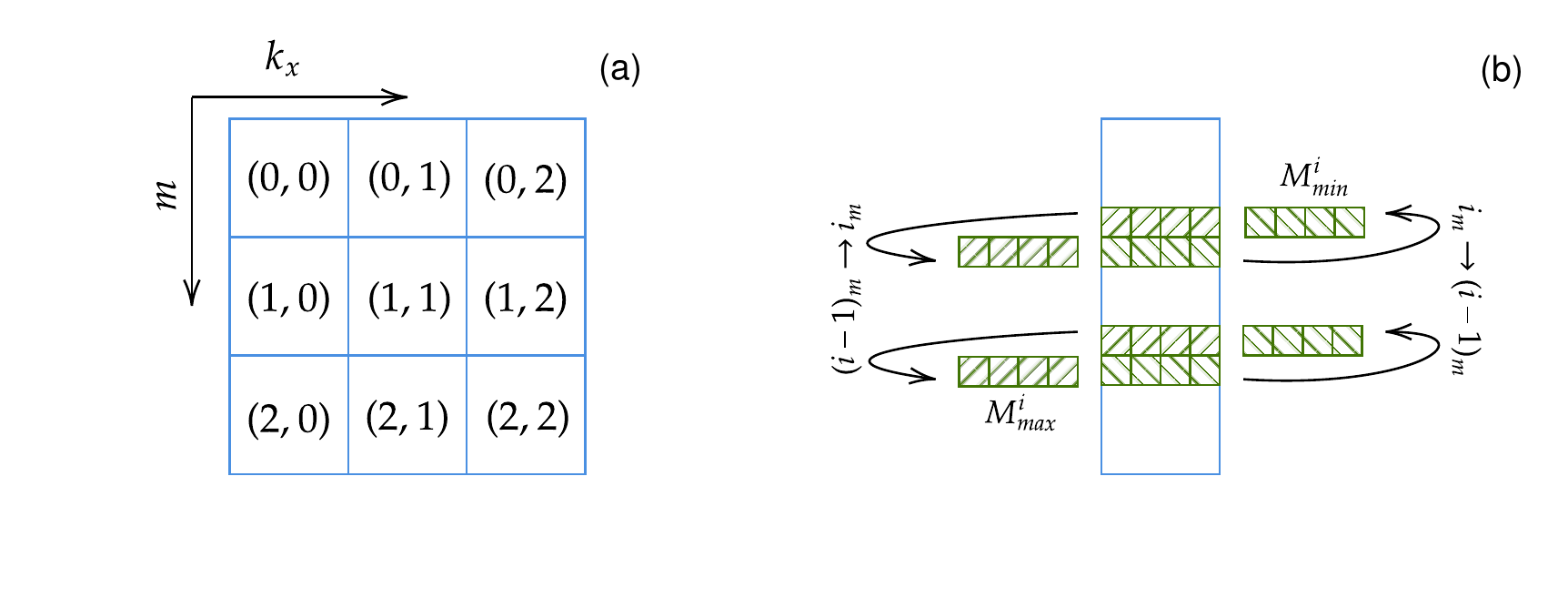}
\caption{(a) two-dimensional Cartesian MPI topology implemented in \code{ALLIANCE}. Hermite moments are distributed along columns, and $k_x$ are distributed along rows. (b) Boundary Hermite moments exchange between processors in one column. For a processor with coordinate $i_m$ largest Hermite moment $M^i_{max}$ is transferred to an $(i+1)_m$ processor, and smallest Hermite moment $M^i_{min}$ is transferred to $(i-1)_m$ processor.}
\label{fig:parallelization}
\end{figure}
{\cc In a typical simulation run, two Laguerre moments are solved at most. This is hardcoded in {\tt ALLIANCE} by the construction of the DK equations \cite{gorbunov_2022} (i.e. truncated at $l > 1$), making the choice of $l$ axis as the direction for the parallelisation inexpedient. However, the amount of Hermite moments $m$ is not limited by the model, as well as the spatial resolution of the simulation box, making those axis
preferable for the parallelization. To keep the parallelisation scheme simple, the data is parallelised only along $k_x$ ($y$ for real arrays) and m directions. In future, parallelisation over species should be considered, as well as $k_y$ direction. The parallelisation is done via creating a two-dimensional topology}, with each processor having its own Cartesian grid coordinates $\{i_x, i_m\}$ (see fig.\ref{fig:parallelization}(a)). To simplify the data transfer, two separate MPI communicators \code{mpi\_kx\_comm} and \code{mpi\_m\_comm} are used, one for the data exchange along the $k_x$ direction and the other one is for the data transfer along the $m$ direction, respectively. The fields and the gyrokinetic potentials are distributed only along $k_x$ direction, and each processor with the same $i_m$ coordinate stores the same $k_x$ part of the fields.

During the simulation run, two operations require the exchange of data between processors. The first one is the computation of the linear term, which requires the exchange of border cells in the Hermite direction. If the processor with coordinates $(i_x, i_m)$ has access to the data slice spanning between the Hermite moments $M^i_{min},M^i_{max}$, then it will send the slice of the distribution function $g(\mathbf{k},M^i_{max},l,s)$ to the processor $\{i_x,(i+1)_m\}$, and $g(\mathbf{k},M^i_{min},l,s)$ to the processor $\{i_x,(i-1)_m\}$, in order to compute the linear term. The same communication is performed when computing the nonlinear term, in order to obtain the contribution to the $m$-space exchange due to $\chi^A$. {\cc Another data exchange is performed in $k_x$ during the FFT, and is treated by {\tt FFTW} library.}

\subsection{Energy injection}\label{sec:Forcing}
When the collision terms are active, if one desires to compensate for the energy dissipation throughout the simulation run, a "forcing" mechanism is needed to inject energy in the system. \code{ALLIANCE} incorporates a simple forcing mechanism, that ensures constant power $\epsilon$ injection throughout the simulation run. First, we define the free energy as in \cite{gorbunov_2022} 
\begin{align}\label{eq:freeEnergyGH}
W = \sum_s T_s n_s \sum_l\sum_k g^m_{l,s}(\mathbf{k}) \left(h^m_{l,s}(\mathbf{k})\right)^*
\end{align}
And we define the forcing operator as following 
\begin{align}
    &\frac{\partial g^m_{l,s}(\mathbf{k})}{\partial t} = \mathcal{F}[h],\label{eq:forcingOperator}\\
    &\mathcal{F}[h] = 
        \begin{cases}
            \alpha h^m_{l,s}(\mathbf{k})\delta_{l,0}\delta_{m,m'}, &\text{if }  k^f_{min}<|k_\perp|\leq k^f_{max}\\\\
            0, & \text{otherwise}\\
        \end{cases}
\end{align}
The forcing of such form only excites zeroth Laguerre moment of the distribution function within the shell $k^f_{min},k^f_{max}$ and the Hermite moment $m'$. The user of the code decides on the forcing shell boundaries, and which Hermite moment he desires to force. We now, however, need to define a parameter $\alpha$, a local forcing amplitude at point $(k_x,k_y,k_z,m)$. To do so, we multiply \eqref{eq:forcingOperator} by $T_s n_s h_{l,s}^{m*}$, sum over all the phase space coordinates and species, and obtain the following equation 
\begin{align}
    \frac{dW(t)}{dt} = \alpha \sum_s T_s n_s \sum_l\sum_m\sum_k |h^m_{l,s}(\mathbf{k})|^2 \delta_{l,0}\delta_{m,m'}
\end{align}
using the relation \eqref{eq:freeEnergyGH}. Now, considering the constant power injection 
\begin{align}
    \frac{d W(t)}{d t} = \epsilon
\end{align}
we obtain the relation for the local forcing amplitude as 
\begin{align}
    \alpha = \epsilon\left[\sum_s T_s n_s \sum_k |h^{m'}_{0,s}(\mathbf{k})|^2\right]^{-1}.
\end{align}
Naturally, different forcing mechanisms can also be implemented by the user.

\subsection{Free energy diagnostics}
\code{ALLIANCE} outputs the full data of the distribution function moments as well as the electromagnetic fields, both in the position and wave space. However, a significant disk space required to store the 6-dimensional array, and the three 3-dimensional arrays makes the frequent output of the raw simulation data unfavourable. Therefore, several reduced-data diagnostics are implemented in \code{ALLIANCE}, based on the analysis of the free energy, and are computed in the real-time during the simulation run. As in works \cite{howes_2006,teaca_2021,Schekochihin_2008}, the free energy can be computed from the gyrokinetic distribution function and the electromagnetic fields as 
\begin{align}
    &W = \sum_\mathbf{k} \left[\sum_m W_h(\mathbf{k},m) - W_\phi(\mathbf{k}) + W_{B_\parallel}(\mathbf{k}) + W_{B_\perp}(\mathbf{k})\right]\,,\label{eq:freeEnergy}
\end{align}
which includes the contributions from the free energy density functions $W_{h}(\mathbf{k},m) = \sum_{s} W_{h,s}(\mathbf{k},m)$ and the fields defined as 
\begin{align}
    &W_{h,s}(\mathbf{k},m) = T_s n_s\sum_{l}\frac{|h^m_{l,s}(\mathbf{k})|^2}{2}\label{eq:W_h}\,,\\
    &W_\phi(\mathbf{k}) =\sum_s \frac{q^2_s n_s |\phi(\mathbf{k})|^2}{2T_s}\label{eq:W_phi}\,,\\
    &W_{B_\parallel}(\mathbf{k}) = \frac{|B_{\parallel}(\mathbf{k})|^2}{8\pi}\label{eq:W_B}\,,\\
    &W_{B_\perp}(\mathbf{k}) = \frac{k^2_{\perp}|A_{\parallel}(\mathbf{k})|^2}{8\pi}\,.\label{eq:W_A}
\end{align}
We define several diagnostics for the free energy. 

\emph{$\bullet$ Free energy channels}: we compute each contribution to the free energy from the free energy spectral-density functions \eqref{eq:W_h}-\eqref{eq:W_A} as $W_h = \sum_\mathbf{k} \sum_m W_h(\mathbf{k},m)$, $W_\phi = \sum_\mathbf{k} W_\phi(\mathbf{k})$, $W_{B_\perp} = \sum_\mathbf{k} W_{B_\perp}$. 

\emph{$\bullet$ Free energy spectra}:
    We also compute spectral quantities of each free energy contribution in $k_\perp$. Each 
    \begin{align}
        &W^{spec}_{\phi, B_\parallel, B_\perp}(k_\perp) = \sum_{\mathbf{k}\in \mathbf{k}_{shell}} \frac{1}{N_{shell}} W_{\phi, B_\parallel, B_\perp}(\mathbf{k}),\\
    \end{align}
    and for for the entropic contribution of the $h^m_{ls}(\mathbf{k})$ to the free energy the two-dimensional spectra is computed in $(k_\perp,m)$ space as
    \begin{align}
        &W^{spec}_{h}(k_\perp,m) = \sum_{s}T_s n_s\sum_{l}\sum_{\mathbf{k}\in \mathbf{k}_{shell}} \frac{1}{N_{shell}} \frac{|h^m_{l,s}(\mathbf{k})|^2}{2},\\
    \end{align}
    and the total free energy spectra is computed as sum over the individual contributions 
    \begin{align}
        &W^{spec}(k_\perp) = \sum_m W^{spec}_{h}(k_\perp,m) - W^{spec}_{\phi}(k_\perp) + W^{spec}_{B_\parallel}(k_\perp) + W^{spec}_{B_\perp}(k_\perp)\\
    \end{align}
    There are two options to the spectral shell initialization in \code{ALLIANCE}. The first one utilizes equidistant shells, with the shell boundaries $k^i_{shell}$ and the shell centres $k^{i}_{center}$ computed as 
    \begin{align}
        &k^i_{shell} = k_0 \cdot i, \quad i\in\{0,N_{shells}\},\\
        &k^i_{center} = \frac{1}{2}\left(k^{i}_{shell} + k^{i+1}_{shell}\right), 
    \end{align}
    The amount of the wave vectors falling inside one shell is then computed for the each shell. For this choice, size of the unit shell $k_0$ is provided by user, and a number of shells $N_{shells}$ is computed automatically from the size of the simulation box. Second choice utilizes the logarithmically spaced shells, each computed using the golden ratio relation
    \begin{align}
        &k^i_{shell} = \varphi^{\frac{i}{2}}\cdot k_0, \quad i\in\{0,N_{shells}\}\\
        &\varphi = \frac{1+\sqrt{5}}{2},\\
    \end{align}
    For this choice of the shell spacing, user also have to provide only size of the unit shell $k_0$.
    
\emph{$\bullet$ Nonlinear flux spectra}:
Nonlinear flux spectra uses the same shell disctretization of the wave space as any other spectral computation in the code. In fact, the nonlinear flux of the free energy through scale $k_c$ is computed as 
    \begin{align}
        \Pi(k_c) = \frac{1}{N_x N_y N_z}\sum_s T_s n_s\sum_{l}\sum_m \sum_\mathbf{k} N^m_{l,s}[h] [{h^{m*}_{l,s}}]^{\le}_{k^c_\perp}(\mathbf{k})
    \end{align}
    The operator $[*]^{\le}_{k^c_\perp}$ denotes sharp low-pass filtering \cite{teaca_2021} over the cut-off wave number $k^c_\perp$. In order to obtain the nonlinear flux, \code{ALLIANCE} computes the nonlinear term only once during the function call, and then performs several computations of $[{h^m_{l,s}}]^{\le}_{k^c_\perp}(\mathbf{k})$. It is should be noted, however, that it is possible to compute the nonlinear flux in the position space as well, as it was done in work \cite{teaca_2021}.
    
\emph{$\bullet$ Dissipative and integral scale lengths}: we compute dissipative and integral length scales as 
\begin{align}
    &L_{int} = 2\pi\frac{\sum_\mathbf{k}1/k_\perp W^{spec}(k_\perp)}{\sum_\mathbf{k} W^{spec}(k_\perp)}\,,\\
    &L_{dis} = 2\pi\frac{\sum_\mathbf{k}k^{2\gamma_\perp - 1}_\perp W^{spec}(k_\perp)}{\sum_\mathbf{k}k^{2\gamma_\perp}_\perp W^{spec}(k_\perp)}\,.
\end{align}
Such estimates allow to form an assumption on the largest length scale of the system and the scale where the dissipation peeks in intensity.

\section{Code validation}\label{sec:CodeValidation}
In this section, we present numerical validation of the code, such as the convergence of the numerical solver. At this stage, we do not engage in the study of any particular physical effects, which will be done elsewhere. However, we show that \code{ALLIANCE} is capable of describing relevant plasma phenomena, such as the linear phase mixing and the turbulent cascade. Necessary tests regarding the free energy conservation are also performed. While the numerical box resolution both in $\mathbf{k}$ and $m$ are varied, several parameters were fixed for the runs: $m_e/m_i = 0.01$, $T_e/T_i = |q_e|/q_i = n_e/n_i = \beta = 1$, $L_x = L_y = 20\pi$, $L_z = 200\pi$. 
\subsection{Energy conservation for linear simulation}
\begin{figure}[h]
\centering 
\includegraphics[width=0.375\textwidth]{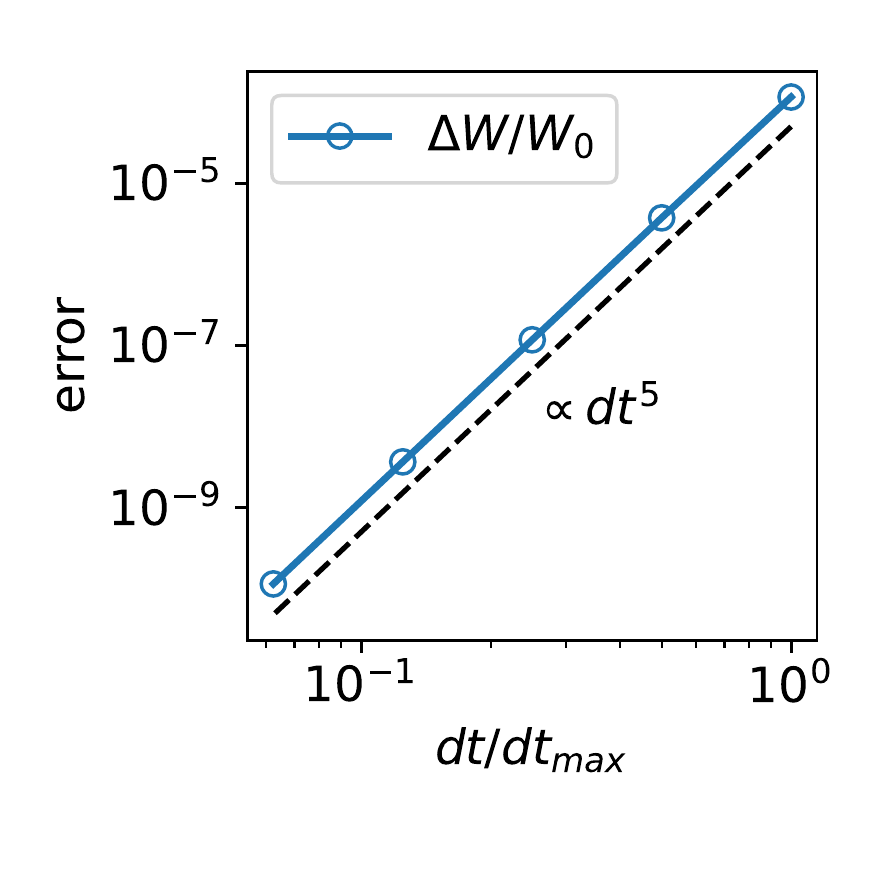}
\caption{Free energy error for varying time step. With decrease of the time step, free energy converges. Scaling of the error is shown by dashed line ($\propto dt^5$) }
\label{fig:linearError}
\end{figure}
In order to check the convergence of the linear solution, several runs for the system with $N_m = 128$ were performed with lack of forcing or collisions. Since for the linear runs there are no nonlinear interactions presented, spatial size of the system is not important, and can be arbitrary. The time step was different for each run, $dt = \{dt_{max}, \frac{dt_{max}}{2}, \frac{dt_{max}}{4}, \frac{dt_{max}}{8},\frac{dt_{max}}{16}\}$, where $dt_{max}$ was chosen as half of the time step size at which RK4 method becomes unstable for the system. Each time, the simulation was run for the total simulation time $t_{total} = 10^3 dt_{max}$, and variance of the free energy, $\Delta W = W_{t_{max}} - W_0$ was measured. Fig.~\ref{fig:linearError} proves the convergence of the RK4 for the linear simulations, with the free energy error scaling as $\propto dt^5$.

\subsection{Nonlinear simulation energy error and adaptive time stepping}

\begin{figure}[h]
\centering 
\includegraphics[width=0.75\textwidth]{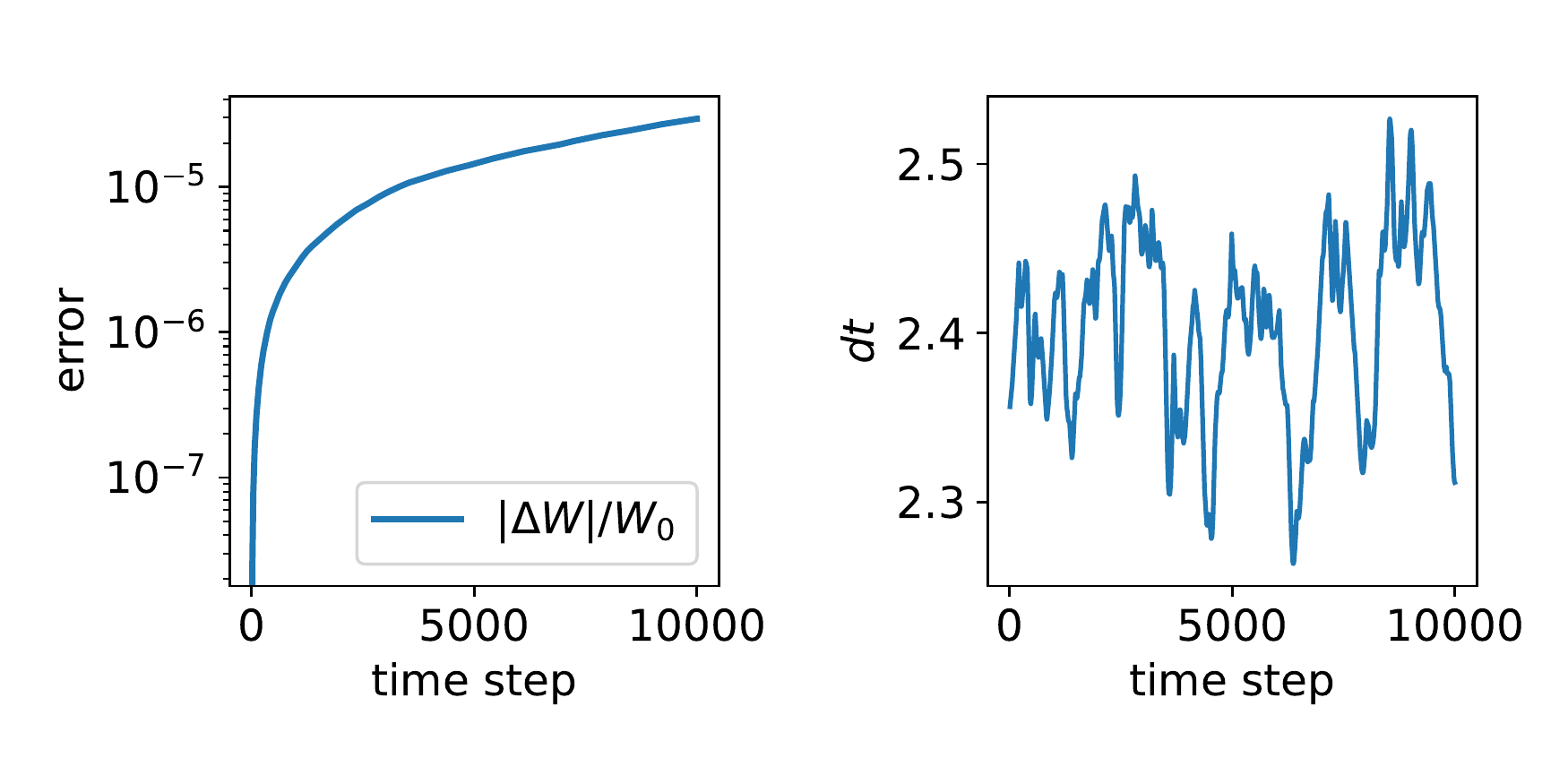}
\caption{(a) Free energy error evolution for the nonlinear run. (b) time step size changes}
\label{fig:nonlinearError}
\end{figure}

To check the nonlinear implementation, we perform a series of purely nonlinear run, where the RHS of the eqs.\eqref{eq:Vlasov0Hermite}-\eqref{eq:Vlasov1Hermite} neglect all other terms (i.e. linear) except \eqref{eq:nonlinearTerm0}-\eqref{eq:nonlinearTerm1}, we have tracked the evolution of the free energy error. The system of size $(N_x,N_y,N_z,N_m) = (64,64,64,2)$ was evolved from the random initial conditions for $10^3$ steps, and the free energy was measured each 10 steps. Fig.\ref{fig:nonlinearError} shows that the the time step ($dt$) is oscillating in a bounded regime, a fact that allows us in practice to update its value (using the algorithm described in Sec.\ref{sec:numericalImplementation}) every so number of time steps, here every 10 iterations. As also shown in fig.\ref{fig:nonlinearError} the fraction of the free energy loss was equal only to ${\Delta W/W_0}\approx10^{-5}$ by the end of the run.

\subsection{Energy balance run}\label{sec:energyBalance}
\begin{figure}[H]
\centering 
\includegraphics[width=1\textwidth]{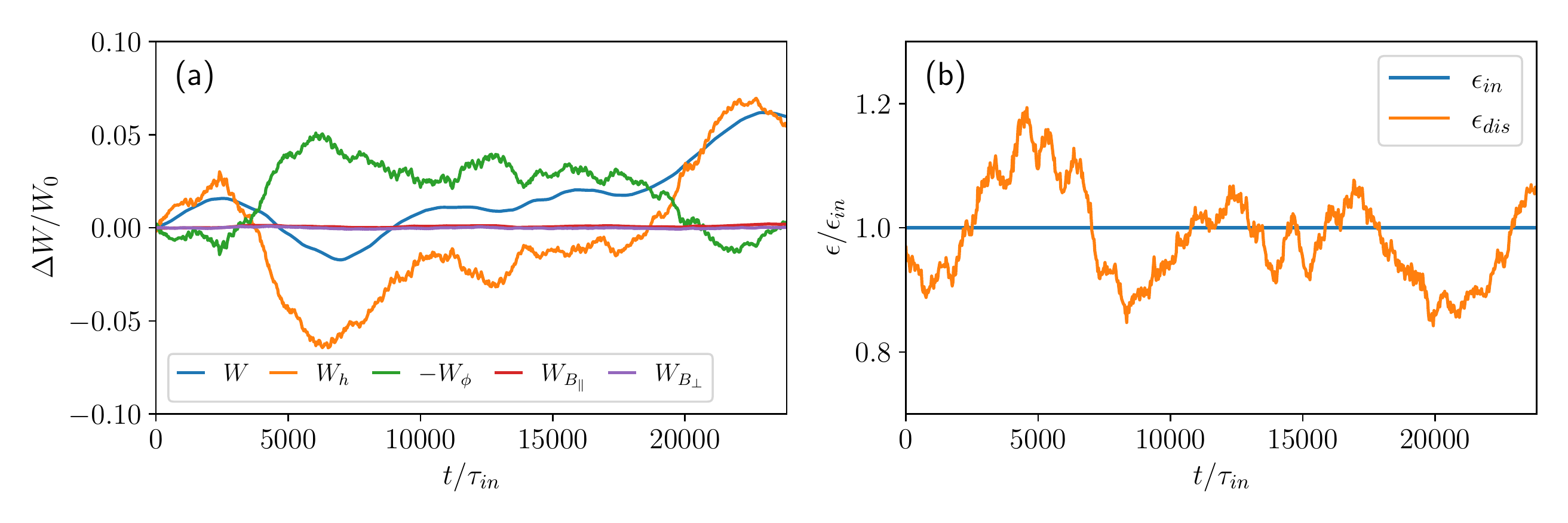}
\caption{(a) The variation in time of the free energy channels. (b) Energy balance: injected power (blue line) and dissipation power (orange line) are shown. For statistical equilibrium, dissipation power value oscillates around the injection power value.}
\label{fig:energyBalance}
\end{figure}
In order to check if the system is capable of reaching the statistical equilibrium with the external forcing provided by \eqref{eq:forcingOperator} and dissipation as \eqref{eq:collisions}, small electromagnetic runs were performed for the system of size $(N_x,N_y,N_z,N_m) = (48,48,48,6)$. The choice of power injection $\epsilon_{in}$ is arbitrary, as in the case of statistical equilibrium dissipation the energy injection should be balanced by collisions. For statistical purposes, we introduce a characteristic injection time, $\tau_{in} = (\frac{2\pi^2 m_i}{k^2_{in}\epsilon_{in}})^{1/3}$, where $k_{in}$ is the wave number at which energy is injected into the system, and verify the equilibrium state over multiples units of $\tau_{in}$. We let the system run until the injected power and the dissipation power are balanced, and measured the variation of the free energy channels, {\cc which took 8 hours on 32 processors. Topology of the parallelisation was chosen to be 16 processors along $k_x$ direction and 2 processors along $m$ direction}. The results of the run are shown in fig.\ref{fig:energyBalance}. The variation of the total free energy is no bigger than $10\%$ in case of the statistical equilibrium, and dissipation power deviates up to $20\%$ from the injection power.

\subsection{Linear phase mixing}
\begin{figure}[h]
\centering 
\includegraphics[width=1\textwidth]{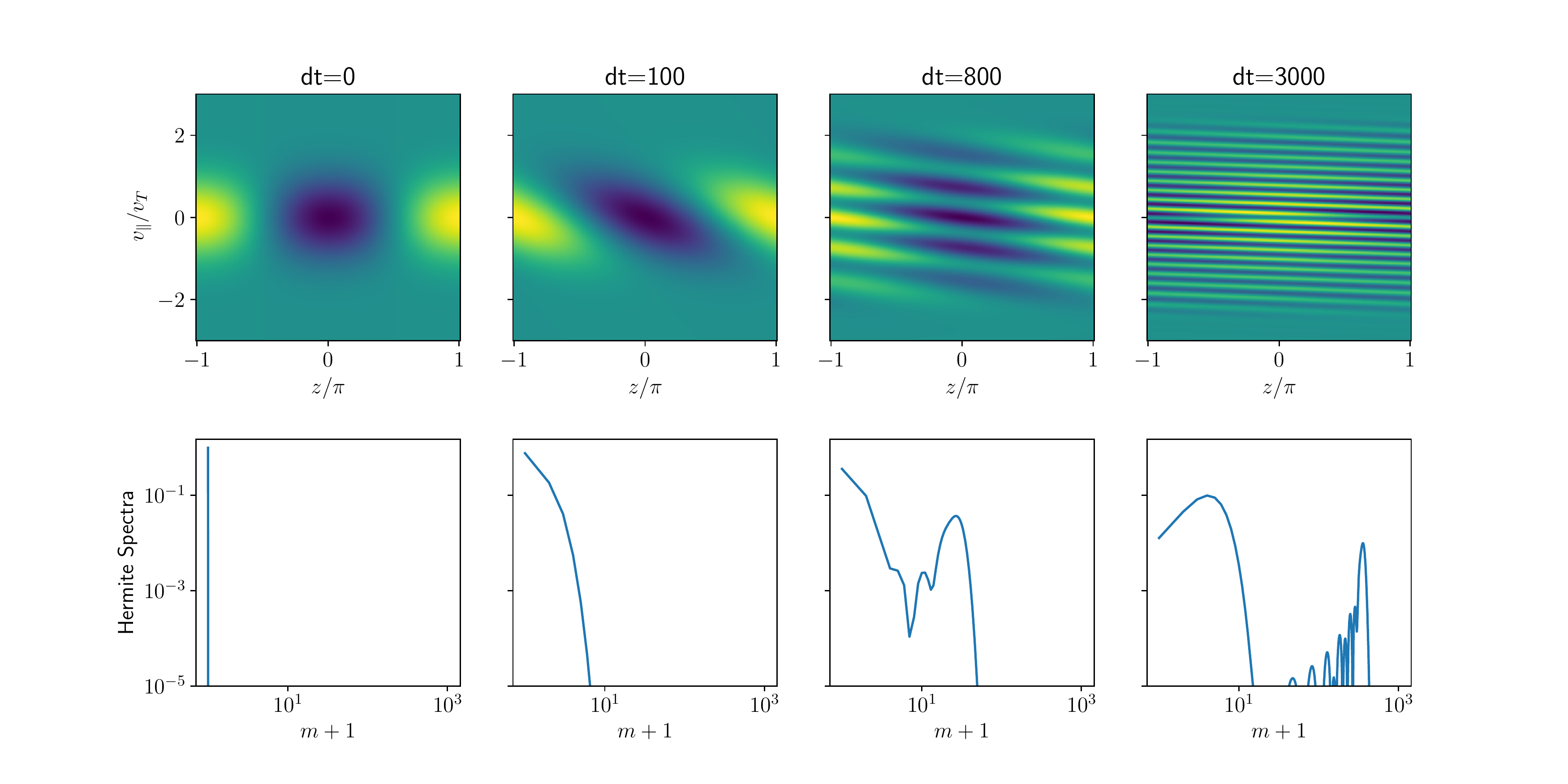}
\caption{{\cc Top panels: linear phase mixing in $(z, v_\parallel)$ direction. Bottom: Hermite spectra evolution. Values smaller than $10^{-5}$ are not shown. The spectra remains non-negative for the whole simulation run.}}
\label{fig:linearMixing}
\end{figure}

To test if the code is capable of capturing linear phase mixing, where a simple $k_z$ spatial perturbation gives rise to finer and finer $v_\|$ structures, we run a simple linear run. We provide a simple initial conditions $h^m_{l,s}(t=0) = q_s cos(z)cos(y)cos(x)\delta_{m0}\delta_{l0}$. Such initial condition gives a simple perturbation in $(z,v_\parallel)$ direction. The $k_x=k_y=1$ mode is exited to overcome the limitation of the gyrokinetic ordering (i.e. $k_\perp<k_\parallel$ is needed, a relation broken when $k_\perp = 0$ is considered). In order to simplify the test as much as possible, we limit ourselves to electrostatic system run $\beta = 0$. {\cc The simulation was performed with $N_m = 1024$. The amount of the Hermite moments does not alter the results. Since no collisions are presented in this run, the large amount of Hermite moments allows for the system to evolve before the recurrence phenomenon \cite{frei2022momentbased} occurs due to finite size of the simulation domain. The total simulation time took around 40 minutes to complete on 16 cores.} We present the evolution of the initial perturbation in fig. \ref{fig:linearMixing}. 

\begin{figure}[H]
\centering 
\includegraphics[width=1\textwidth]{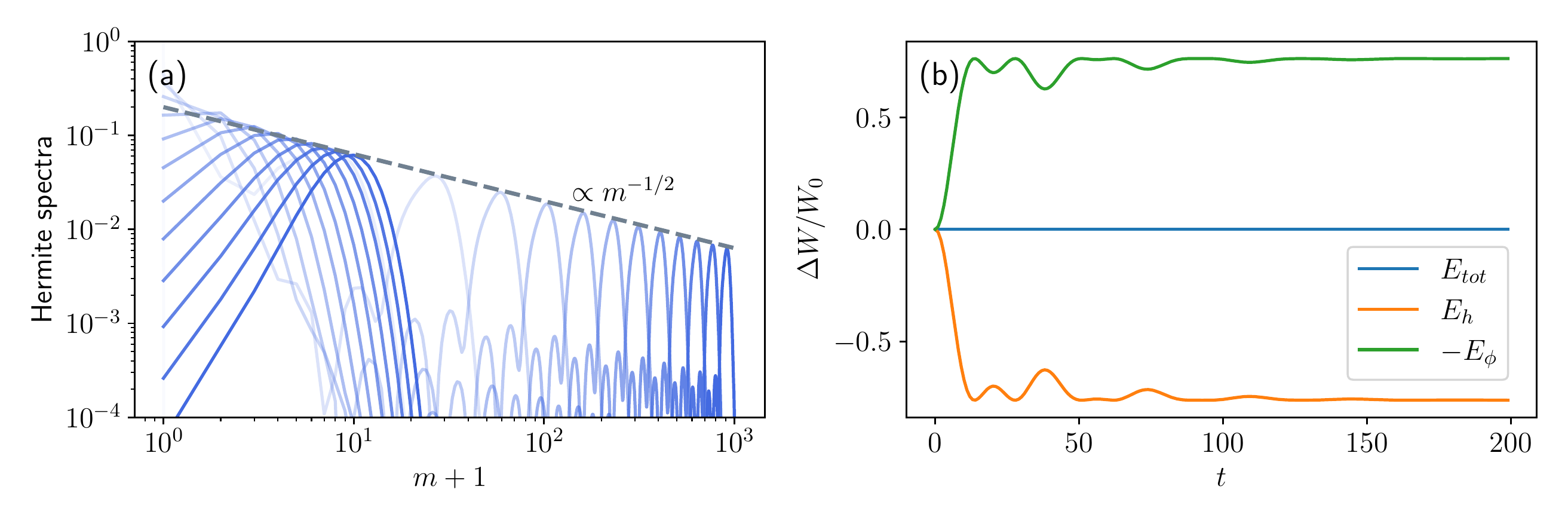}
\caption{(a) Evolution of the initial condition perturbation in Hermite space. {Darker lines are corresponding to later simulation times.} (b) Free energy evolution in case of linear phase mixing.}
\label{fig:linearMixingEnergy}
\end{figure}

The evolution of the Hermite spectra over time and the evolution of the different energy channels is shown in fig.~\ref{fig:linearMixingEnergy}. We observe how the free energy is cascaded in $m$-space with a $m^{-1/2}$ slope, as expected of pure linear mixing. The run is stopped here before the finite size of the system is reached, and reflections of the energy from the largest $m$ to the smallest becomes an issue. As expected, the electrostatic energy $W_{\phi}$ is exchanged with the $W_{h}$, while to total free energy remains constant, as the damping of electrostatic fluctuations are producing fine structure in the $v_\|$ direction.

\begin{figure}[H]
\centering 
\includegraphics[width=0.9\textwidth]{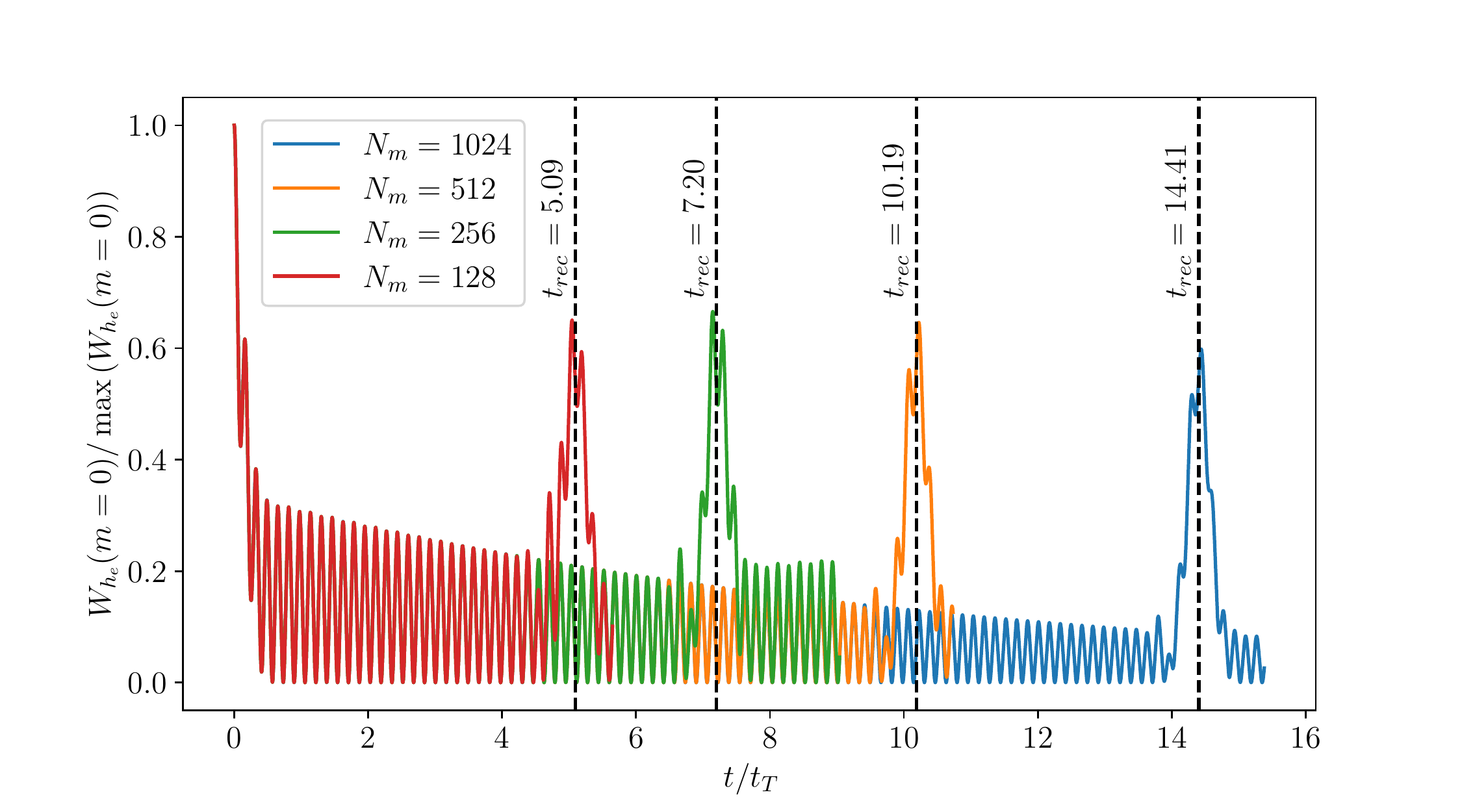}
\caption{{\cc Evolution of the amplitude of the entropic energy \eqref{eq:W_h} for electrons at $m=0$, for systems of sizes $N_m = \{128,256,512,1024\}$ . At $t_{rec} = 2\sqrt{2 N_m}/k_\parallel v_{T_e}$, depicted by the vertical dashed lines, the recurrence occurs. The times are normalised by $t_T = 2\pi/k_\parallel v_{T_e}$.}}
\label{fig:recurrence}
\end{figure}

{\cc For the same simulation set up, we have also studied the dependence of the recurrence time $t_{rec}$ on the amount of Hermite moments $N_m$. As shown in \cite{frei2022momentbased}, due to the finite system size in Hermite space, one can expect the initial perturbation to reflect from the boundary and return back, leading to the recurrence phenomenon, occuring at the time estimated as $t_{rec} = 2\sqrt{2 N_m}/k_\parallel v_{T_s}$. We have observed the change in the amplitude of entropic energy \eqref{eq:W_h} at zeroth Hermite moment $W_h(m=0)$, given in Fig.\ref{fig:recurrence}. At the recurrence time $t_{rec}$, the sharp amplitude increase is observed, as the initial perturbation returns from higher moments to lower ones. The times at which recurrence occurs agrees perfectly with theory \cite{frei2022momentbased}.}

{\cc
\subsection{Linear dispersion relation}

In order to check the linear physics of Alfv\'en waves, the we have performed simple similar run in the similar manner as it was done in \cite{numata_2010}. First, Alfv\'en wave was driven with parallel antenna current $A_{ant}$, excited at wave vector $\mathbf{k}_0 = (k_{\perp 0},k_{\parallel 0})$ with frequency $\omega_0$ and the amplitude $A_0$,
\begin{align}
    & A_{ant} = A_0e^{- i \omega_0 t - i\mathbf{k}_0\cdot\mathbf{r}}. 
\end{align}
The antenna potential modifies Ampere laws \eqref{eq:A},\eqref{eq:numerics:A} as following
\begin{align}
    & A_\parallel(\mathbf{k}) + A_{ant}\delta_{\mathbf{k},\mathbf{k}_0} = \frac{\beta}{2 k_\perp^2}\sum_s  q_s n_s v_{T_s} \sqrt{\frac{1}{2}} \mathcal{J}_{s00}h^1_{0,s},\\ 
    & A_\parallel(\mathbf{k}) = \frac{\beta}{2} \frac{\sum_s q_s n_s v_{Ts}\mathcal{J}_{s00}\sqrt{\frac{1}{2}}g_{0,s}^1}{ k_\perp^2+\frac{\beta}{4}\sum_s \frac{q_s^2 n_s v_{Ts}^2}{T_s}\mathcal{J}_{s00}^2} - \frac{A_{ant}\delta_{\mathbf{k},\mathbf{k}_0}}{1 +\frac{\beta}{4 k_\perp^2}\sum_s \frac{q_s^2 n_s v_{Ts}^2}{T_s}\mathcal{J}_{s00}^2}. 
\end{align}
Eq.\eqref{eq:Vlasov0Hermite} takes the form 
\begin{align}
    &\frac{\partial{{g}}^m_{0,s}}{\partial t} = \mathcal{L}_{0,s}^m[h] + \sqrt{\frac{1}{2}}\frac{\partial \chi_{ant}}{\partial t}\delta_{m,1}, 
\end{align}
where we defined gyroaveraged gyrokinetic antenna potential as 
\begin{align}
    &\chi_{ant} = -v_{T_s} J_{s00} A_{ant}. 
\end{align}
In order to measure $A_\parallel$ response to antenna excitation, we fix $k_{\parallel 0} = 1$ and set the antenna to be slightly off-resonant from the Alfv\'en frequency $\omega_A$ as $\omega_0 = 0.9 \omega_A$, and excited the Alfv\'en wave at different $k_{\perp 0}$ in range  $0.1<k_{\perp 0}\rho_i<1$. By fitting $A_\parallel$ field response with Fourier-Laplace solution for gyrokinetics \cite{numata_2010}, we were able to find frequencies $\omega = \omega_r + i \gamma$ of Alfv\'en waves. For these runs, the parallel component of the magnetic field was forced to be zero, $B_\parallel = 0$, realistic mass ratio used $m_i/m_e \approx 1836$, $T_i/T_e=n_i/n_e = 1$, and $\beta = 1$. In order to get rid of effects arising due to recurrence phenomenon, a small dissipation in $m$ direction was added. For these simulations we used $N_m = 256$ Hermite moments. The typical simulation run took around 10 minutes to complete. In fig.\ref{fig:KAW}, we show the linear dispersion results obtained, along with theoretically predicted dispersion relation for gyrokinetics (GK) and drift kinetics (DK). For the drift kinetics, the linear dispersion relation can be obtained in the same manner as it was done for GK \cite{howes_2006}, replacing the integrals in $\mu$ by appropriate approximations of Bessel function moments used in this work (see \ref{appendix:dispersion} for details). The measured results start to diverge slightly from the theoretical predictions at higher wave numbers, due to high-frequency oscillations. As it can be seen, {\tt ALLIANCE} is capable of capturing linear KAW physics correctly.
\begin{figure}[h]
\centering 
\includegraphics[width=0.9\textwidth]{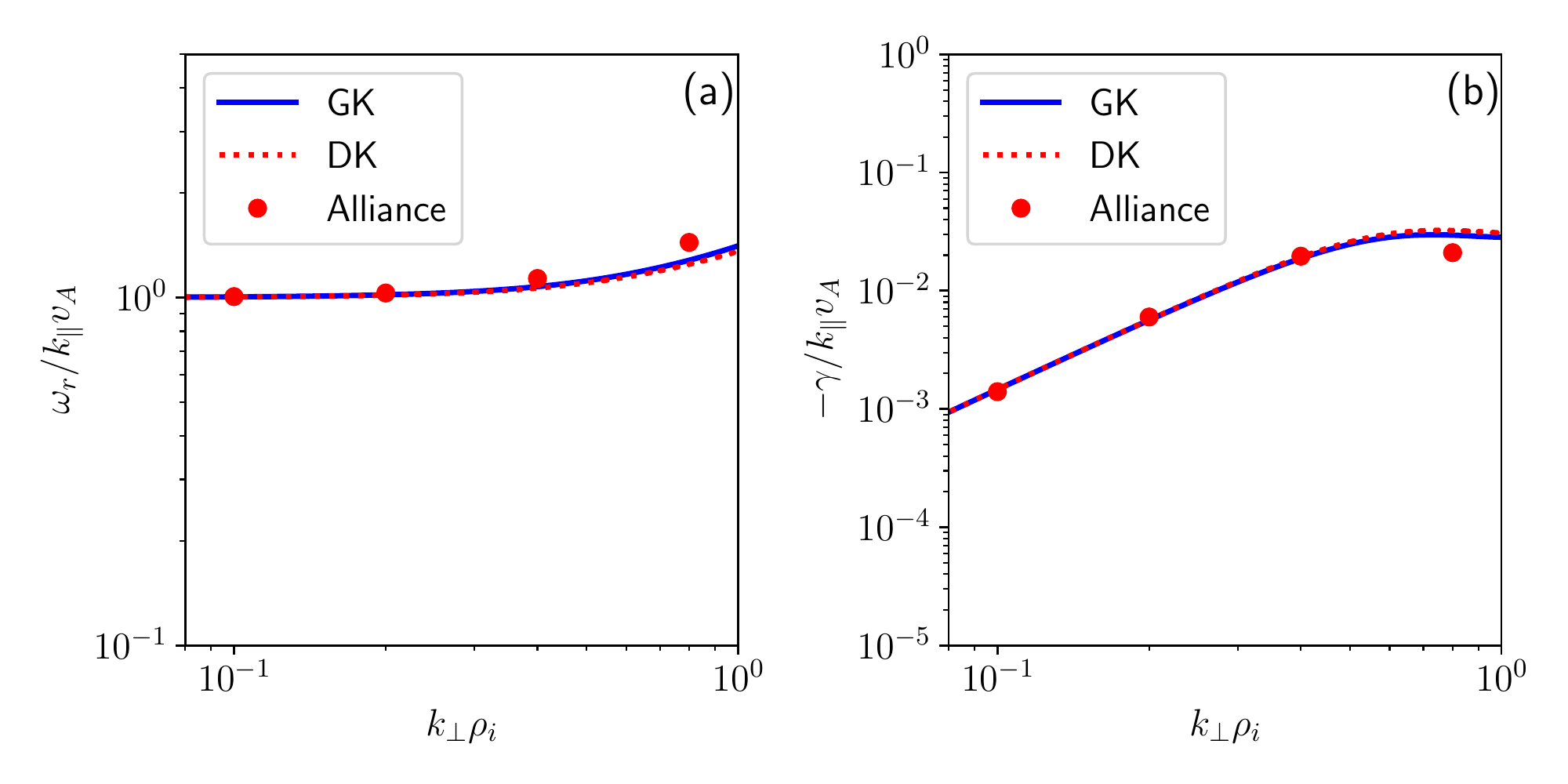}
\caption{(a) Real frequency  and (b) damping rate, obtained analytically from gyrokinetic dispersion relation (blue solid line), drift kinetic equations (dotted red line), and measured from {\tt ALLIANCE} (red dots). At higher wave numbers the difference between the analytical results and measurements arise due to existence of higher frequency oscillations.}
\label{fig:KAW}
\end{figure}
}
\subsection{Nonlinear run}

\begin{figure}[h]
\centering 
\includegraphics[width=0.9\textwidth]{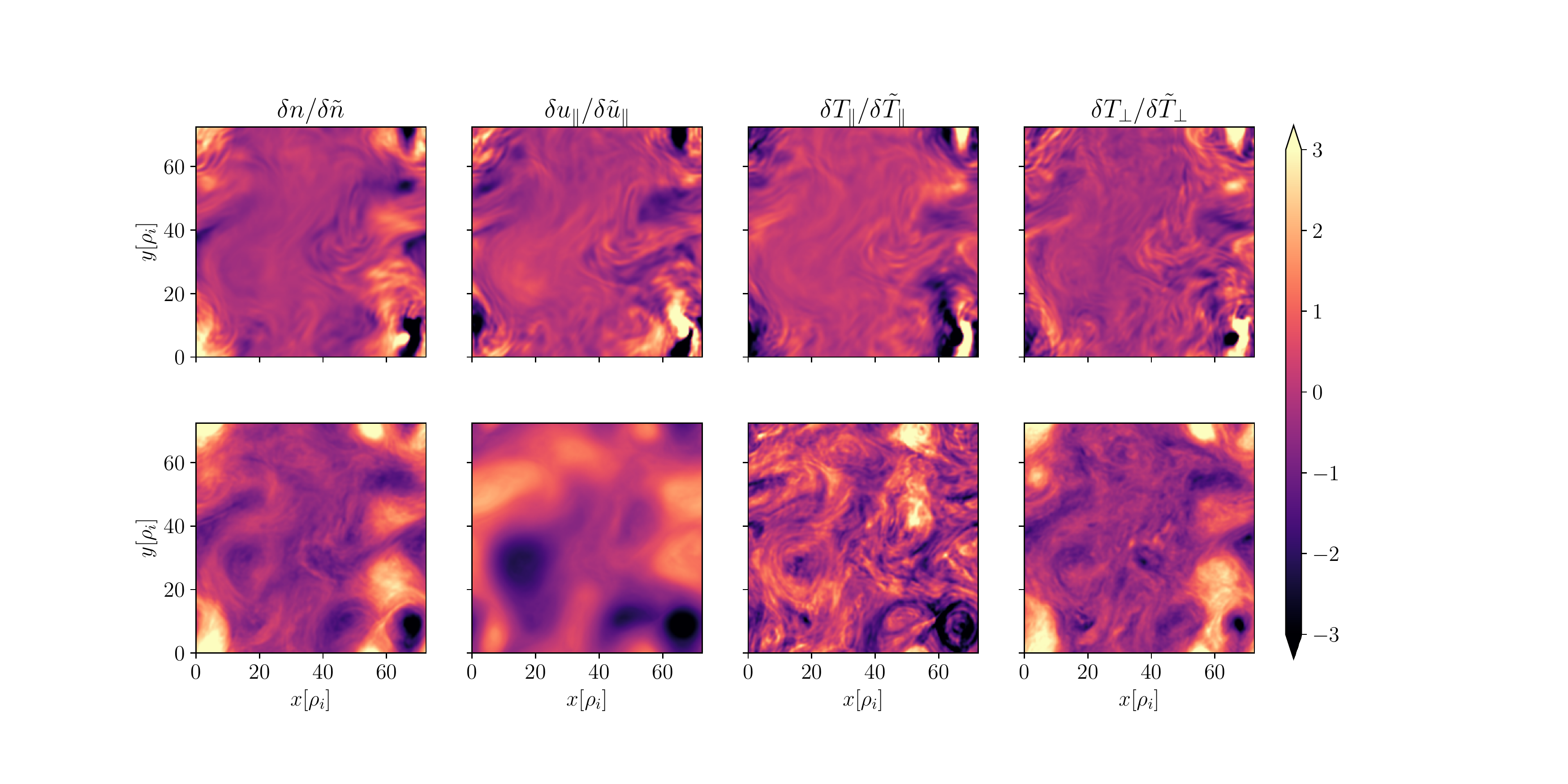}
\caption{Gyrocenter distribution function moments, slice in z, normalized by their standard deviation. Top row: ions, bottom row: electrons.}
\label{fig:moments}
\end{figure}
We present here results from a higher resolution nonlinear run, $(N_x,N_y,N_z,N_m) = (192,192,192,32)$, initialised from the steady state run described in \ref{sec:energyBalance}. The system was evolved for $t = 300 \tau_{in}$, until the beginning of statistical steady state was reached. However, the typical properties of the developed turbulence can already be observed. {\cc The total time to for the system reach the state shown in this section was 3 days on 128 processors.}

\begin{figure}[H]
\centering 
\includegraphics[width=0.6\textwidth]{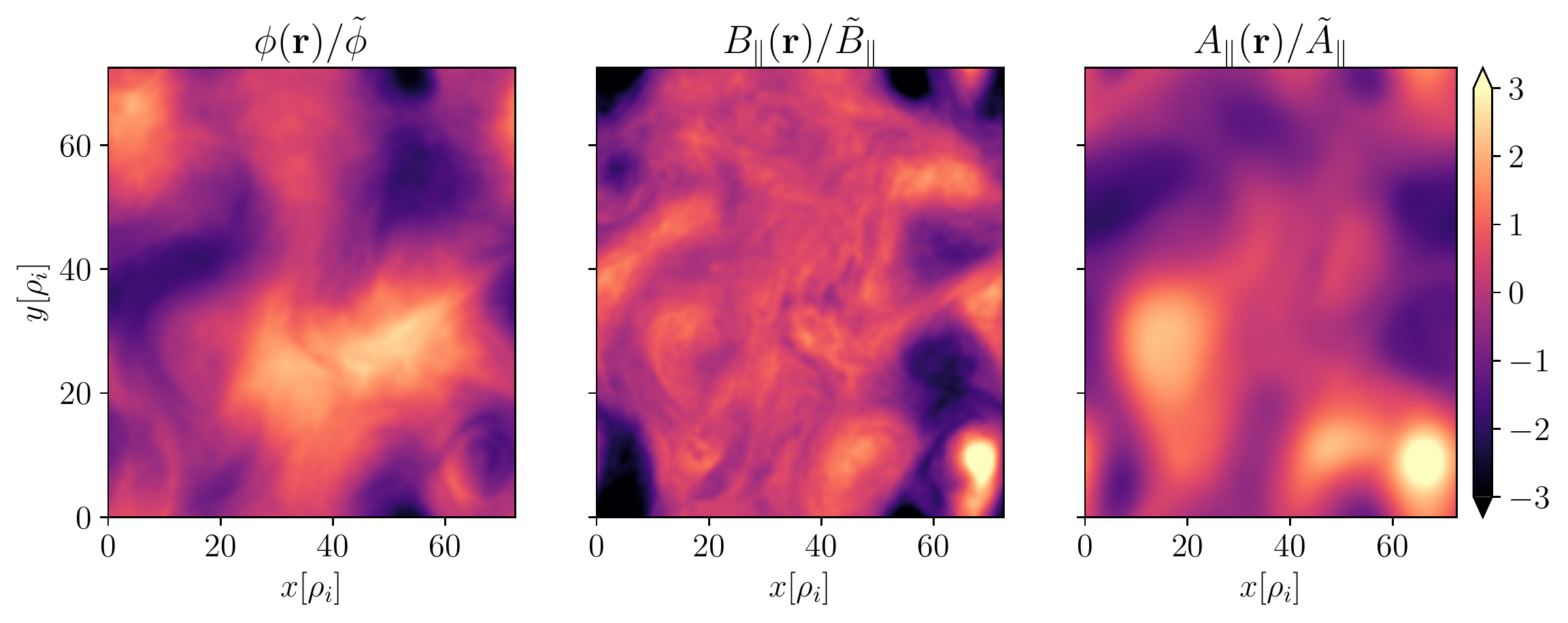}
\caption{slice in z for electromagnetic fields, normalized by their standard deviation. $B_\parallel$ develops finest structures among the fields, and $A_\parallel$, tends to develop largest structures, as all the finest are being damped by $1/k^2_\perp$ term in \eqref{eq:numerics:A}.}
\label{fig:fields}
\end{figure}

First, we present the space density of the first moments of the gyrocenter distribution function, which are being studied in many gyro-fluid models \cite{Beer1996,tassi_2020}. Considering that all the normalization have been applied, the relation between the gyrofluid moments and the gyrocenter distribution function Hermite-Laguerre moments are $(\delta n_s, \delta u_{\parallel,s}, T_{\parallel,s}, \delta T_{\perp,s}) = (g^0_{0,s},\frac{1}{\sqrt{2}} g^1_{0,s},\sqrt{2} g^2_{0,s},g^0_{1,s})$, with $\delta n_s$ being the gyrocenter density perturbation, $\delta u_\parallel$ is the parallel velocity perturbation, and $\delta T_\perp, \delta T_\parallel$ are the perpendicular and parallel temperature perturbations, respectively. These quantities  are shown in fig. \ref{fig:moments} both for the ions (top row) and electrons (bottom row), normalized by their standard deviations at a given $z$-plane, i.e. for an arbitrary function $f$: 
\begin{align}
    &\tilde{f} = \sqrt{\frac{1}{N_x N_y}\sum_{i_x = 0}^{N_x}\sum_{i_y = 0}^{N_y}\left[f(i_x,i_y,z)-\langle f \rangle\right]^2}\,,\\
    &\langle f \rangle = \frac{1}{N_x N_y}\sum_{i_x = 0}^{N_x}\sum_{i_y = 0}^{N_y} f(i_x,i_y,z).
\end{align}
In addition, we also show the electromagnetic fields in fig.\ref{fig:fields}.

\begin{figure}[t]
\centering 
\includegraphics[width=0.6\textwidth]{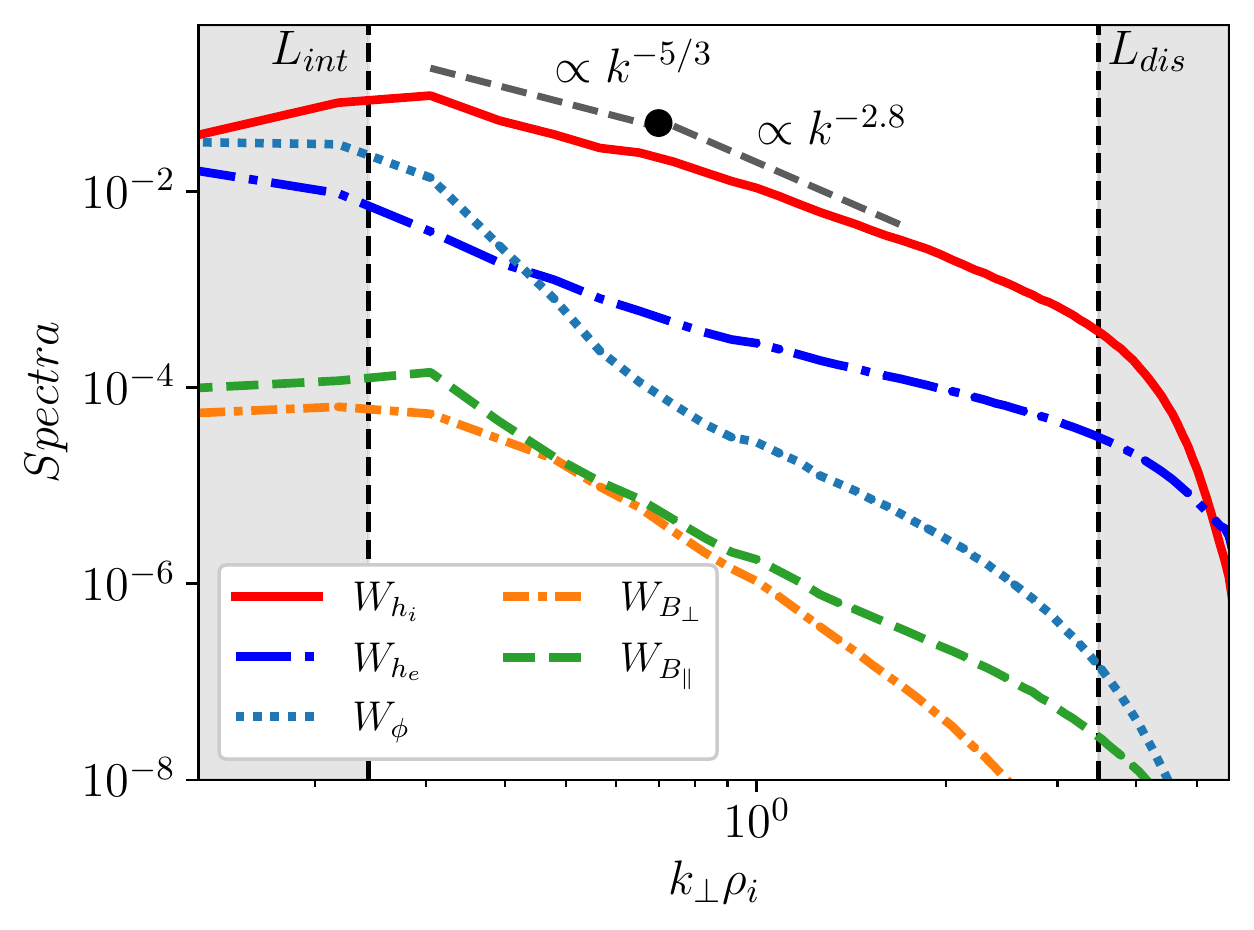}
\caption{Spectra for different channels contributing to total free energy $W$. The black dashed lines are slopes given to reference two regimes: fluid $-5/3$ and gyrokinetic \cite{told_2015} $-2.8$.}
\label{fig:cascade}
\end{figure}

One of the properties of the turbulence is the existence of the cascade in the so-called inertial range, i.e. the wavenumber range between the large integral scale where energy is injected and the dissipation scale where the energy is thermalized by collisions or other mechanisms. Please note that for a kinetic plasma, the inertial range definition is changed to includes the possibility of linear phase mixing (velocity mixing) taking place. In order to check if what is observed in the simulation is indeed turbulence, in fig.\ref{fig:cascade}, we plot the spectra of different channels of the free energy, including the contributions of the electromagnetic fields \eqref{eq:W_phi}-\eqref{eq:W_A} and the entropic contributions \eqref{eq:W_h}. Not only that we see a power-law spectrum specific of turbulence, we also see a tendency for the spectral exponent to change around $k_\perp \rho_i \sim 1$, specific for magnetized plasma. While encouraging, the physics of the spectral-break should be carefully investigated for runs that employing realistic $m_e/m_i$ mass ratios, and which use much larger resolutions to allow for a better scale separation.

\begin{figure}[h]
\centering 
\includegraphics[width=0.65\textwidth]{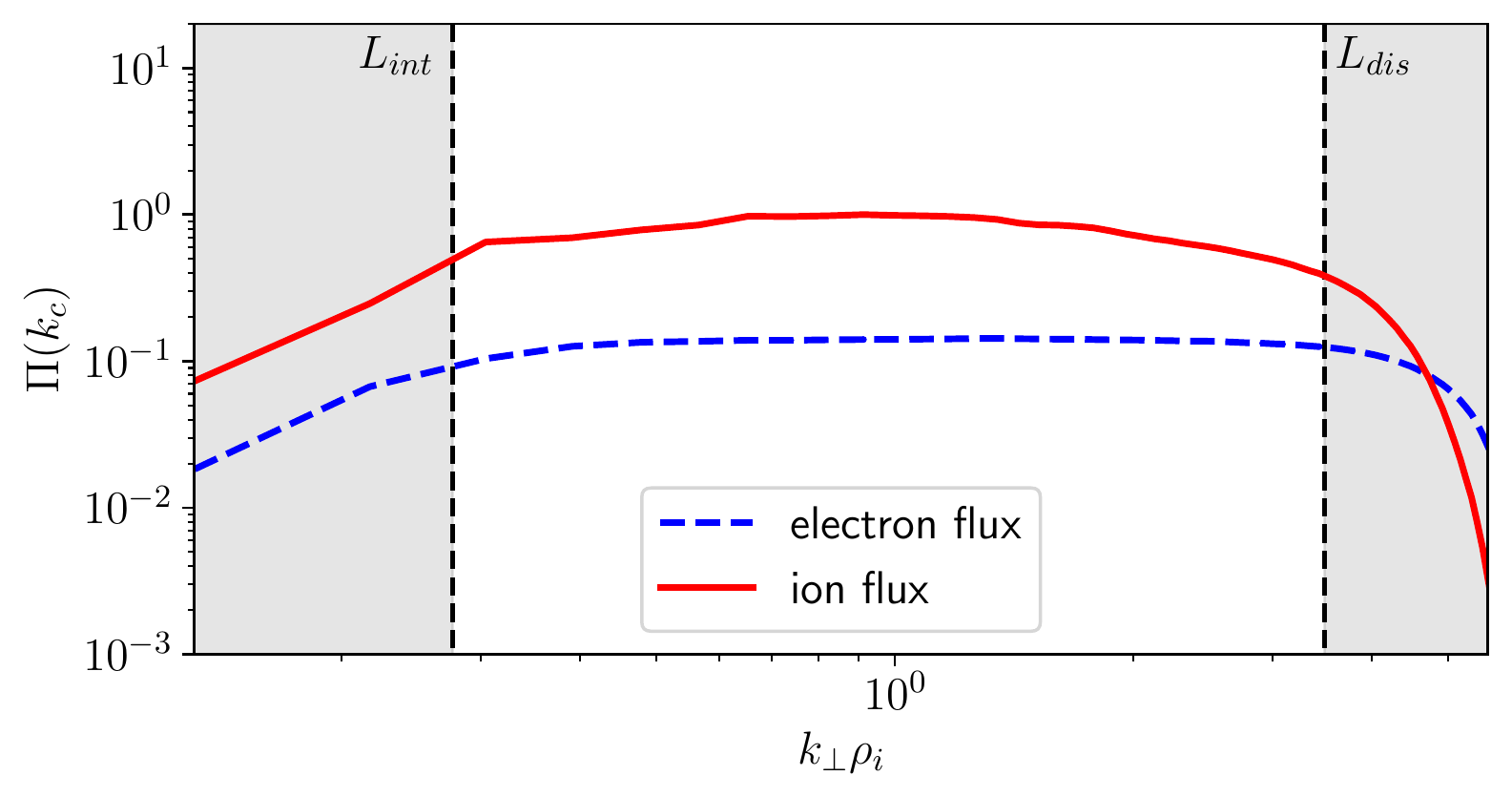}
\caption{Nonlinear flux for ions  and electrons. }
\label{fig:flux}
\end{figure}

In fig. \ref{fig:flux}, looking at the wavenumber flux of free-energy for the two species, we see better that the electrons are exhibiting a scale-invariant cascade typical for fluid turbulence, while ions exhibit a change in the nature of the cascade around $k_\perp \rho_i \sim 1$.

\begin{figure}[H]
\centering 
\includegraphics[width=0.55\textwidth]{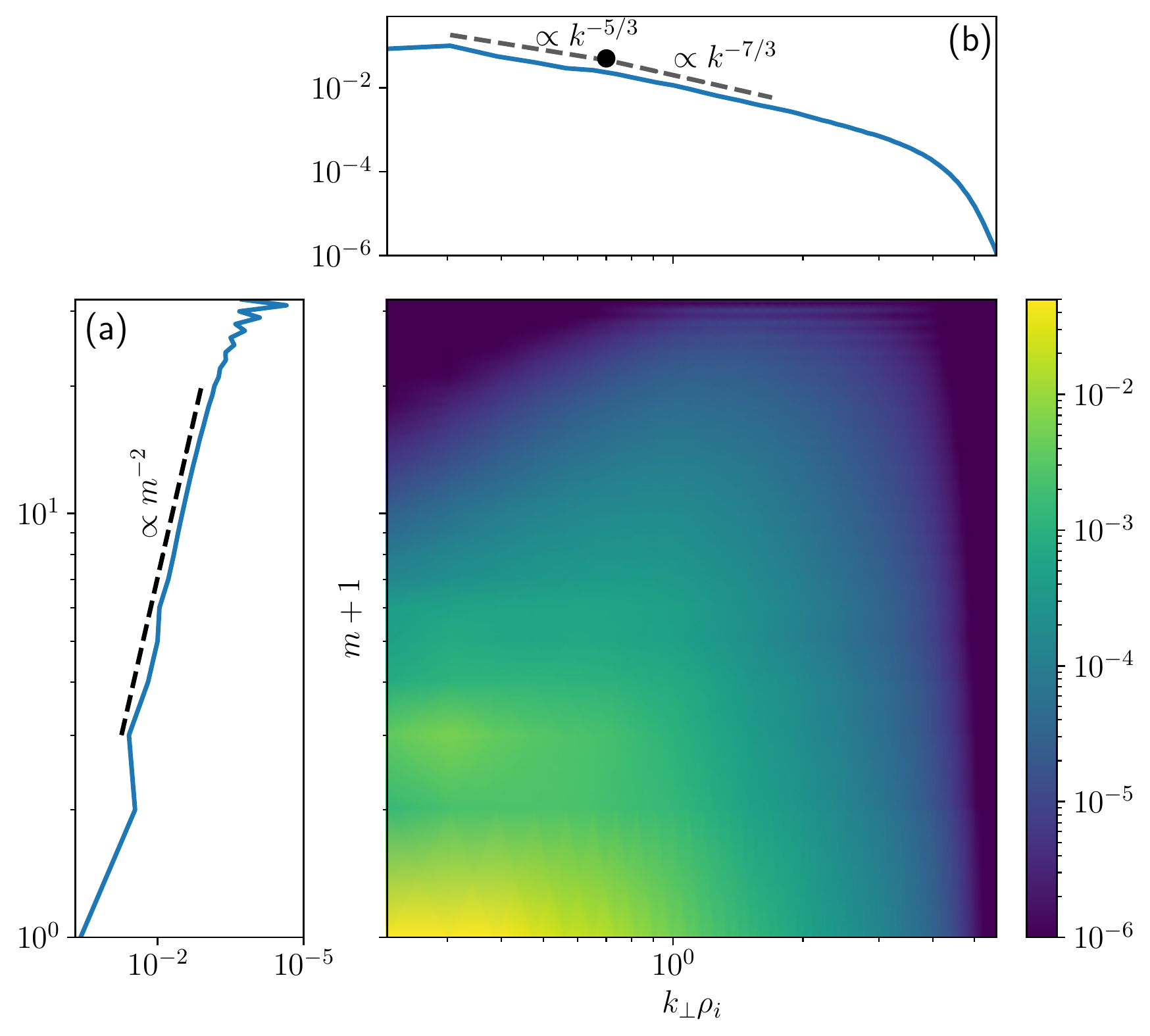}
\caption{Fourier-Hermite $W_{h,i}(k_\perp,m)$ spectra. (a) $W_{h,i}(k_\perp,m)$ spectra integrated over $k_\perp$. (b) $W_{h,i}(k_\perp,m)$ spectra integrated over m.}
\label{fig:2dSpeci}
\end{figure}

\begin{figure}[H]
\centering 
\includegraphics[width=0.55\textwidth]{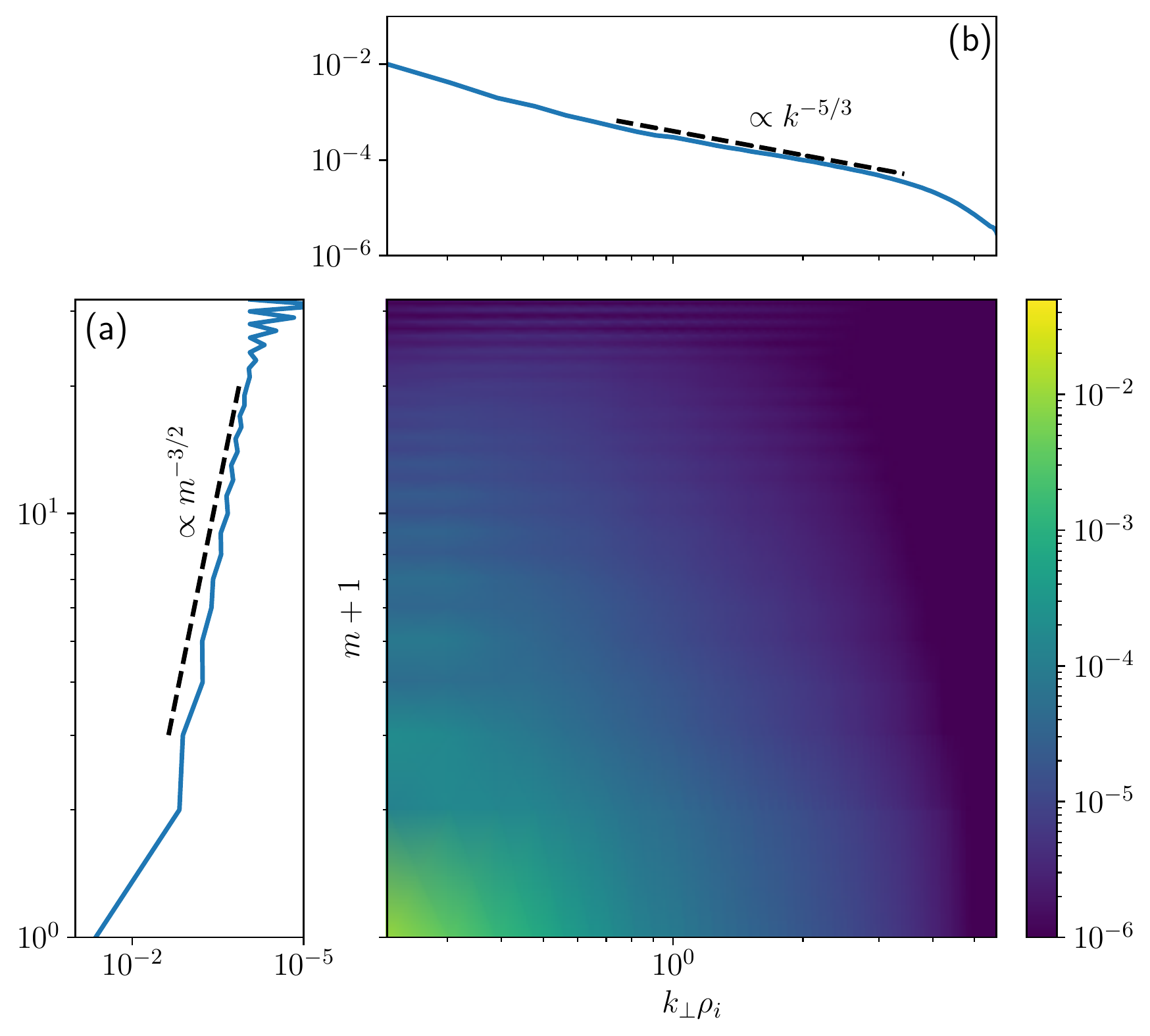}
\caption{Fourier-Hermite $W_{h,e}(k_\perp,m)$ spectra. (a) $W_{h,e}(k_\perp,m)$ spectra integrated over $k_\perp$. (b) $W_{h,e}(k_\perp,m)$ spectra integrated over m.}
\label{fig:2dSpece}
\end{figure}

Apart from the cascade in the $k_\perp$ direction, the drift-kinetic model also allows to study the cascade in the parallel velocity direction. We show the two-dimensional $(k_\perp,m)$ spectra of \eqref{eq:W_h} in fig.\ref{fig:2dSpeci} for the ions and in fig.\ref{fig:2dSpece} for the electrons. The integrated Hermite spectra, as well as $k_\perp$ one, are shown on the insets to that figure. The presence of linear-mixing is evident from these pictures.

\section{Scalability}\label{sec:PerformanceRuns}
In this section, we investigate the performance of the \code{ALLIANCE} code. We measure the time required for the solver to integrate one step (time per step) for a single processor, as well as perform weak and strong scaling tests to measure a parallel performance. All the numerical tests were performed with the physical parameters same as in sec.\ref{sec:CodeValidation}. The energy is injected into the systems during the runs, and  the dissipation is also included. 

\subsection{Single processor scalability}

\begin{figure}[h]
\centering 
\includegraphics[width=0.75\textwidth]{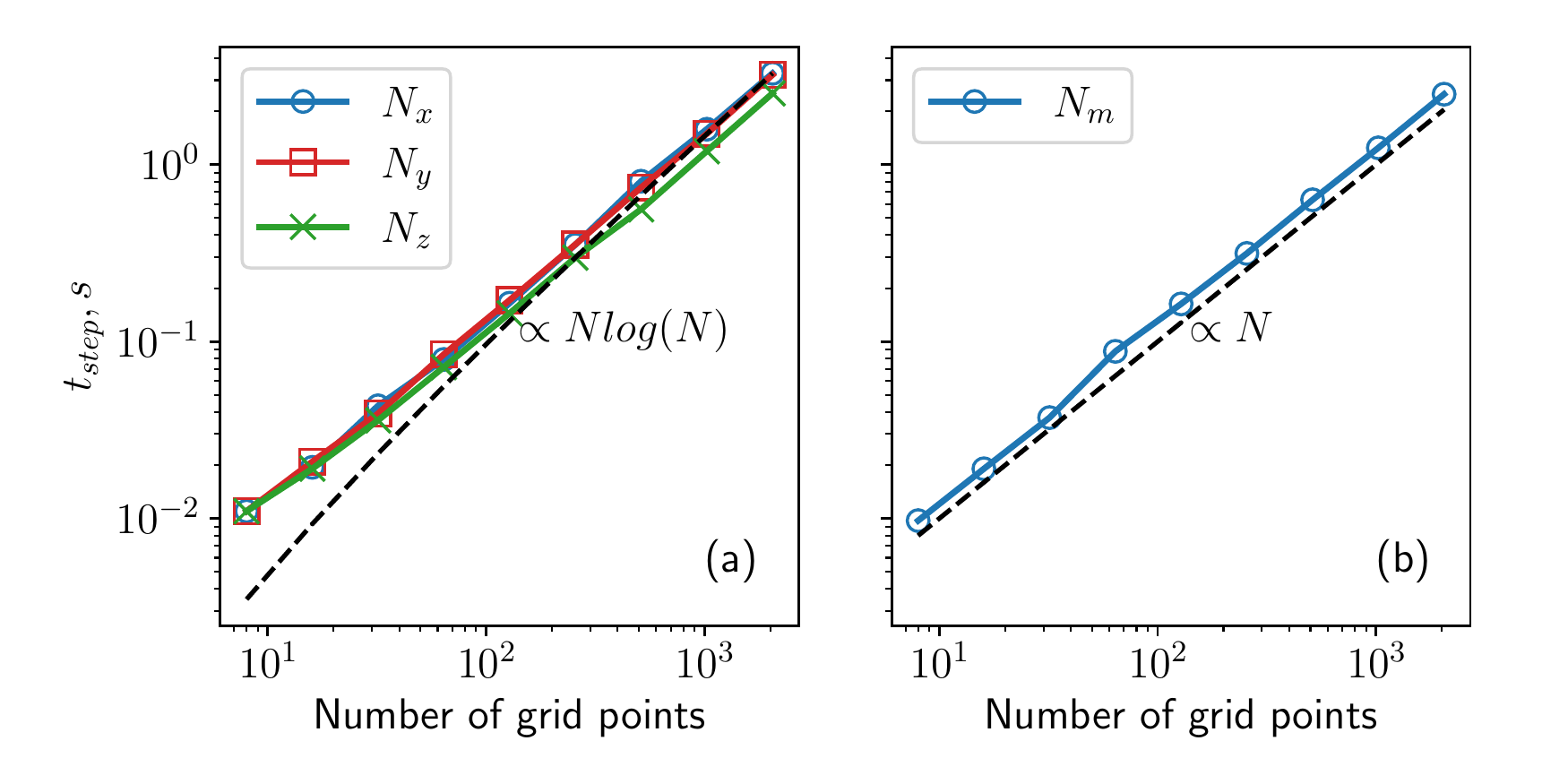}
\caption{Time for a single solver step depending on the number of grid points. (a) Time per step variation for different amount of grid points in $x,y,z$-directions. $N \log(N)$ scaling is asymptotically achieved. (b) Time per step variation for different amount of Hermite moments. Linear scaling ($\propto N$) is achieved.  }
\label{fig:singleProc}
\end{figure}

We tested the scaling of the time per step with increase of different dimensions of the system. Tests were performed on a single Intel Xeon Gold 6140 2.30 Hz processor in order to eliminate the communication time between processors. The initial system size for this test was $(N_x,N_y,N_z,N_m) = (8,8,8,8)$. For each run, one of the dimensions was increased by 2 up to 2048 grid points. For each run the time per step was measured. The results of the tests are shown in fig.\ref{fig:singleProc}. As expected, for $(x,y,z)$-dimensions the time per step scales as $N \log N$, asymptotic behaviour of the fast Fourier transform. Since the increase of the grid points in Hermite direction increases the problem size linearly, the time per step also growth in a linear fashion. Therefore, we expect for time per step to scale as $t_{step} \propto (N_x \log N_x)(N_y \log N_y)(N_z \log N_z)(N_m)$.

\subsection{Parallel scalability}

\begin{figure}[h]
\centering 
\includegraphics[width=0.75\textwidth]{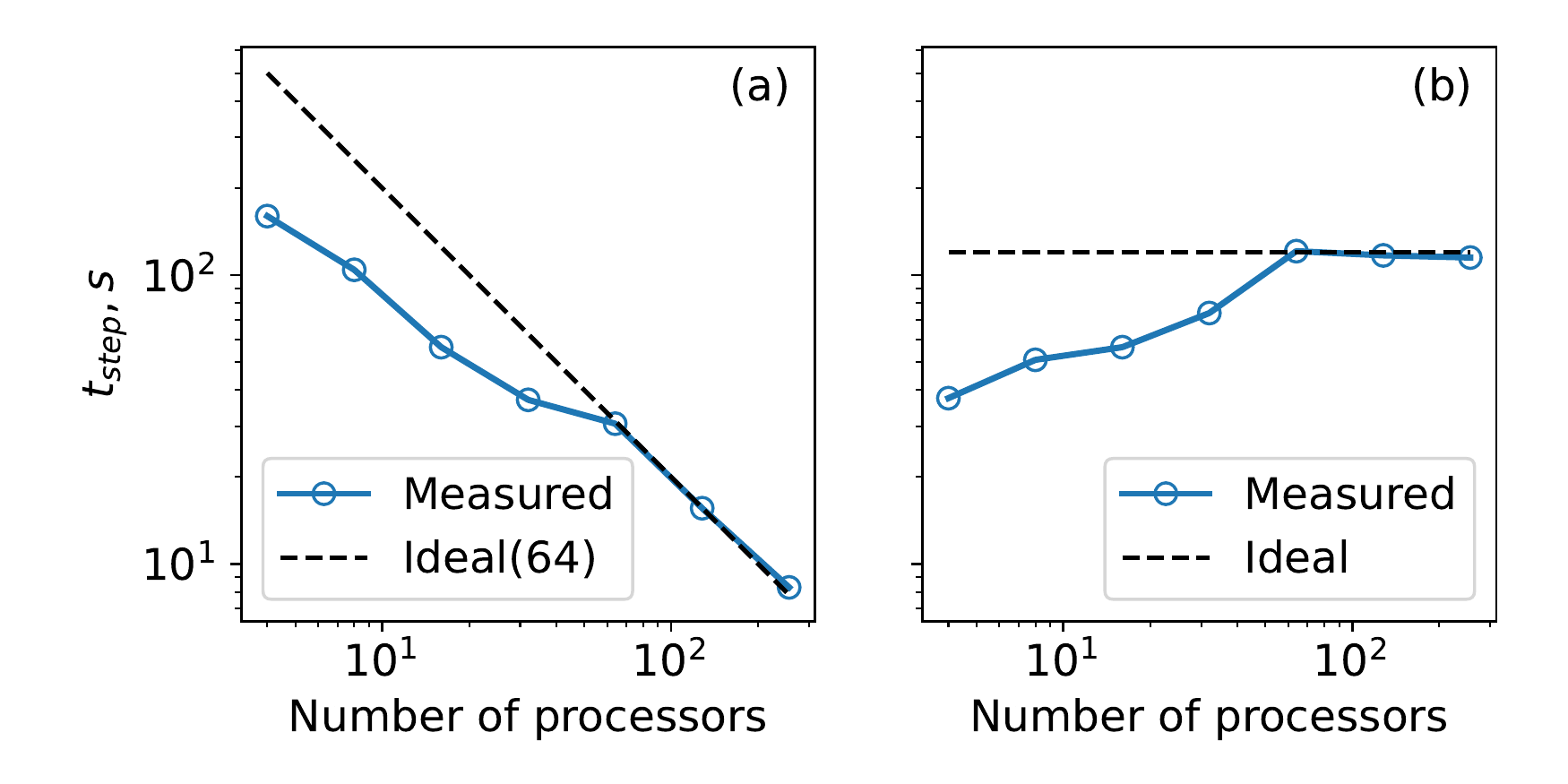}
\caption{(a) Strong scaling test. The time per step starts scaling ideally from 64 to 256 processors, as demonstrated by the ideal linear scaling for 64 processors. (b) Weak scaling test. The time per step does not increase from 64 to 256 processors.}
\label{fig:multiProc}
\end{figure}

The parallel performance of \code{ALLIANCE} was measured with the two standard tests, weak scaling and strong scaling. Both tests were performed at Coventry's University EPYC cluster consisting of AMD EPYC 7742 64-core processors. For these tests, the number of processors was varied from 4 to 256. 

In case of the strong scaling, the size of the simulation remains constant, while the number of processors is being increased. Initial problem had the size of $(N_x,N_y,N_z,N_m) = (128,128,128,128)$. Every time the processor amount was doubled, either $k_x$ or $m$ direction of parallelization was distributed, interchangeably. Measured wall-clock time per step of the simulation is shown in fig.\ref{fig:parallelization}(a). Performance degrades up to 64 processors, showing almost ideal scaling from 64 to 256 processors. {\cc We believe that the ideal scaling is achieved due to inter-processor communication, since each EPYC processor has 64 cores. Different system configuration did not show significant change in the behaviour of the weak and strong scaling.}

For the weak scaling, the single processor size of the problem was chosen to be $(N_x,N_y,N_z,N_m) = (32,128,128,32)$. The $k_x$ and $m$ dimensions were doubled interchangeably for this test, with results shown in fig.\ref{fig:parallelization}(b). As for the strong scaling, performance degrades up to 64 processors due to increase of inter-processor communications, but from 64 up to 256 there is no significant change in the wall-clock time. 

\section{Conclusions}\label{sec:Conclusion}
In this paper, the new pseudo-spectral code \code{ALLIANCE} was presented. The code integrates numerically the new set of the drift-kinetic equations \cite{gorbunov_2022}. Several diagnostics that are implemented have been presented. To facilitate the study of the turbulent cascade, the simple forcing and dissipation mechanisms that are implemented have been presented. The structure of the code, however, allows for an easy modification of the form of those mechanisms, as needed. Initial test runs, both linear and nonlinear were performed, showing the ability of the  model to capture the linear phase mixing and allow for turbulence development.

To solve the equations, \code{ALLIANCE} uses explicit Runge-Kutta of order 4 scheme, allowing excellent accuracy for both linear and nonlinear terms. For the linear simulations, the error scales as $\propto dt^5$. For the nonlinear term, the CFL condition is employed to ensure numerical stability of the nonlinear simulations. 

The algorithm demonstrates ideal single processor scaling, showing expected asymptotic behaviour with increase of the system size for all the dimensions. \code{ALLIANCE} also demonstrates great parallel performance, for which further optimizations will be considered. Moreover, possible further developments of the \code{ALLIANCE} {\cc include transferring of the core functions on GPU similar to other modern codes \cite{mandell2022gx}}, as well as using pencil domain decomposition for FFT transforms in the future to allow better scalability at large supercomputing clusters ($>10^3$ processors). One of the advantages of the \code{ALLIANCE} is, however, that it is being highly portable, with ability to run on a large range of machines, from personal computers to high-performance supercomputing clusters. 

The equations used at the core of \code{ALLIANCE} are specifically designed to be a link between kinetic and fluid description for plasma, thus allowing to study different physical phenomena. The pseudo-spectral approach used along with Hermite decomposition of the parallel velocity direction allows to study kinetic turbulent cascade and its interplay with linear phenomena such as Landau damping, and it can facilitate fundamental studies similar to \cite{Teaca2012, told_2015, Navarro_2016, Teaca_2017, Teaca_2019}, which can offer insight into the fundamental nonlinear structure of magnetized plasma turbulence, especially when a specific scale (i.e. $k_\perp \rho_i = 1$) has a privileged nonlinear impact. The \code{ALLIANCE} code can also help to shed light on the knee break in the solar wind spectra, it can be used to study how large-scale dynamics of plasma affects micro scales, or how fluid constraints at large scales impact the fluid-kinetic turbulent transition of plasma (seen as the point when velocity space dynamics become important).

Apart from this, the connection between gyrofluid moments and the gyrokinetic distribution moments provided here allows to use \code{ALLIANCE} to model gyrofluid system, provided with appropriate closures for the system. Thus, using \code{ALLIANCE}, it is possible to study simpler systems, when full kinetic description is unnecessary to use. It is, therefore, possible to study 3D reconnection using the code, providing possibility to further develop current advances on the topic \cite{granier_2022}.

{\cc
\appendix
\section{Change to the GK linear dispersion relation for the drift kinetic model}\label{appendix:dispersion}
To obtain the linear dispersion relation for the drift kinetic formalism modelled by {\tt ALLIANCE}, it is enough to replace integrals in $\mu$ space (see eq.(C9) in \cite{howes_2006}):
\begin{align}
    & \Gamma_0(b_s) \approx J_{s00}^2(b_s), \nonumber\\
    & \Gamma_1(b_s) \approx \tilde{J}_{s10}(b_s) J_{s00}(b_s), \nonumber\\
    & \Gamma_2(b_s) \approx 2\Gamma(b_s).\label{eq:Gamma}
\end{align}
The resulting dispersion relation remains unchanged apart from the integrals \eqref{eq:Gamma}. 
}
\bibliographystyle{elsarticle-num} 
\bibliography{refs}

\end{document}